\documentclass[twocolumn]{aastex62}

\newcommand\black{\color{black}}

\usepackage{amsmath}
\usepackage{multirow}
\usepackage{placeins}

\graphicspath{{./}{figures/}}

\received{??}
\revised{??}
\accepted{??}

\submitjournal{ApJ}

\shorttitle{Inferred dust grain alignment properties in Ophiuchus}
\shortauthors{Pattle et al.}

\begin{document}

\title{JCMT BISTRO Survey observations of the Ophiuchus Molecular Cloud: Dust grain alignment properties inferred using a Ricean noise model}

\correspondingauthor{Kate Pattle}
\email{kpattle@gapp.nthu.edu.tw}
\author[0000-0002-8557-3582]{Kate Pattle}
\affil{Institute for Astronomy and Department of Physics, National Tsing Hua University, Hsinchu 30013, Taiwan}
\affil{National Astronomical Observatory of Japan, National Institutes of Natural Sciences, Osawa, Mitaka, Tokyo 181-8588, Japan}

\author{Shih-Ping Lai}
\affil{Institute for Astronomy and Department of Physics, National Tsing Hua University, Hsinchu 30013, Taiwan}
\affil{Academia Sinica Institute of Astronomy and Astrophysics, P.O. Box 23-141, Taipei 10617, Taiwan}

\author{Tetsuo Hasegawa}
\affil{National Astronomical Observatory of Japan, National Institutes of Natural Sciences, Osawa, Mitaka, Tokyo 181-8588, Japan}

\author{Jia-Wei Wang}
\affil{Institute for Astronomy and Department of Physics, National Tsing Hua University, Hsinchu 30013, Taiwan}

\author{Ray Furuya}
\affil{Institute of Liberal Arts and Sciences, Tokushima University, Minami Jousanajima-machi 1-1, Tokushima 770-8502, Japan}

\author{Derek Ward-Thompson}
\affil{Jeremiah Horrocks Institute, University of Central Lancashire, Preston PR1 2HE, UK}

\author{Pierre Bastien}
\affil{Centre de recherche en astrophysique du Qu\'{e}bec \& d\'{e}partement de physique, Universit\'{e} de Montr\'{e}al, C.P. 6128 Succ. Centre-ville, Montr\'{e}al, QC, H3C 3J7, Canada }

\author{Simon Coud\'{e}}
\affil{SOFIA Science Center, Universities Space Research Association, NASA Ames Research Center, Moffett Field, California 94035, USA}

\author{Chakali Eswaraiah}
\affil{CAS Key Laboratory of FAST, National Astronomical Observatories, Chinese Academy of Sciences, Peopleʼs Republic of China }

\author{Lapo Fanciullo}
\affil{Academia Sinica Institute of Astronomy and Astrophysics, P.O. Box 23-141, Taipei 10617, Taiwan}

\author{James di Francesco}
\affil{NRC Herzberg Astronomy and Astrophysics, 5071 West Saanich Road, Victoria, BC V9E 2E7, Canada}
\affil{Department of Physics and Astronomy, University of Victoria, Victoria, BC V8W 2Y2, Canada}

\author{Thiem Hoang}
\affil{Korea Astronomy and Space Science Institute (KASI), 776 Daedeokdae-ro, Yuseong-gu, Daejeon 34055, Republic of Korea}
\affil{University of Science and Technology, Korea (UST), 217 Gajeong-ro, Yuseong-gu, Daejeon 34113, Republic of Korea}

\author{Gwanjeong Kim}
\affil{Nobeyama Radio Observatory, National Astronomical Observatory of Japan, National Institutes of Natural Sciences, Nobeyama, Minamimaki, Minamisaku, Nagano 384-1305, Japan}

\author{Woojin Kwon}
\affil{Korea Astronomy and Space Science Institute (KASI), 776 Daedeokdae-ro, Yuseong-gu, Daejeon 34055, Republic of Korea}
\affil{University of Science and Technology, Korea (UST), 217 Gajeong-ro, Yuseong-gu, Daejeon 34113, Republic of Korea}

\author{Chang Won Lee}
\affil{Korea Astronomy and Space Science Institute (KASI), 776 Daedeokdae-ro, Yuseong-gu, Daejeon 34055, Republic of Korea}
\affil{University of Science and Technology, Korea (UST), 217 Gajeong-ro, Yuseong-gu, Daejeon 34113, Republic of Korea}

\author{Sheng-Yuan Liu}
\affil{Academia Sinica Institute of Astronomy and Astrophysics, P.O. Box 23-141, Taipei 10617, Taiwan}

\author{Tie Liu}
\affil{Korea Astronomy and Space Science Institute (KASI), 776 Daedeokdae-ro, Yuseong-gu, Daejeon 34055, Republic of Korea}
\affil{East Asian Observatory, 660 N. A'oh\={o}k\={u} Place, University Park, Hilo, Hawaii 96720, USA}

\author{Masafumi Matsumura}
\affil{Faculty of Education and Center for Educational Development and Support, Kagawa University, Saiwai-cho 1-1, Takamatsu, Kagawa, 760-8522, Japan} 

\author{Takashi Onaka}
\affil{Department of Astronomy, Graduate School of Science, The University of Tokyo, 7-3-1 Hongo, Bunkyo-ku, Tokyo 113-0033, Japan}

\author{Sarah Sadavoy}
\affil{Harvard-Smithsonian Center for Astrophysics, 60 Garden Street, Cambridge, MA 02138, USA}

\author{Archana Soam}
\affil{SOFIA Science Center, Universities Space Research Association, NASA Ames Research Center, Moffett Field, California 94035, USA}

\begin{abstract}

The dependence of polarization fraction $p$ on total intensity $I$ in polarized submillimeter emission measurements is typically parameterized as $p\propto I^{-\alpha}$ $(\alpha \leq 1)$, and used to infer dust grain alignment efficiency in star-forming regions, with an index $\alpha=1$ indicating near-total lack of alignment of grains with the magnetic field.  In this work we demonstrate that the non-Gaussian noise characteristics of polarization fraction may produce apparent measurements of $\alpha \sim 1$ even in data with significant signal-to-noise in Stokes $Q$, $U$ and $I$ emission, and so with robust measurements of polarization angle.  We present a simple model demonstrating this behavior, and propose a criterion by which well-characterized measurements of polarization fraction may be identified.  We demonstrate that where our model is applicable, $\alpha$ can be recovered by fitting the $p-I$ relationship with the mean of the Rice distribution, without statistical debiasing of polarization fraction.  We apply our model to JCMT BISTRO Survey POL-2 850$\mu$m observations of three clumps in the Ophiuchus Molecular Cloud, finding that in the externally-illuminated Oph A region, $\alpha\approx 0.34$, while in the more isolated Oph B and C, despite their differing star formation histories, $\alpha \sim 0.6-0.7$.  Our results thus suggest that dust grain alignment in dense gas is more strongly influenced by incident interstellar radiation field than by star formation history.  We further find that grains may remain aligned with the magnetic field at significantly higher gas densities than has previously been believed, thus allowing investigation of magnetic field properties within star-forming clumps and cores.

\end{abstract}


\section{Introduction} \label{sec:intro}

The role of magnetic fields in the star formation process is not well-understood (e.g. \citealt{crutcher2012}).  Observations of magnetic field morphologies in the densest parts of molecular clouds are typically performed indirectly, through submillimeter dust polarization observations (e.g. \citealt{matthews2009}).  In low-density, well-illuminated environments in the interstellar medium (ISM), dust grains are expected to be aligned with their minor axes parallel to the local magnetic field direction \citep{davis1951}.  However, at sufficiently high optical depths, grains are expected to become less efficiently aligned with the magnetic field \citep{andersson2015}.  Recent results have suggested that this occurs at a visual extinction $A_{V}\sim 20-30$ magnitudes (\citealt{whittet2008}; \citealt{alves2014}; \citealt{jones2015}).  This inferred break in behaviour is, in each case so far reported, coincident with a change in tracer from optical/near-infrared extinction polarimetry to submillimeter dust emission polarimetry (\citealt{jones2015}; \citealt{andersson2015}).  Submillimeter emission polarimetry is the only effective wide-area tracer of ISM polarization in dense molecular clouds ($A_{V}\gtrsim 30$ mag) where stars are forming.  It is thus vital to studies of the role of magnetic fields in star formation to know which gas densities are being traced by dust polarization observations.

A commonly-used method of assessing the alignment of grains is to determine the relationship between polarization efficiency and visual extinction \citep{whittet2008,alves2014,jones2015,jones2016}.  In submillimeter studies this is generally treated as a relationship between polarization fraction $p$ and total submillimeter intensity $I$, as polarization efficiency is identical to polarization fraction for optically thin emission \citep{alves2015}, and optically thin submillimeter total intensity is proportional to visual extinction $A_{V}$ for a given temperature (c.f. \citealt{jones2015}; \citealt{santos2017}).  Recent comparisons between $p$ and submillimeter dust opacity measurements (an alternative proxy for $A_{V}$) suggest that the standard assumption in polarization studies of $I\propto A_{V}$ is likely to be too simplistic \citep{juvela2018}.  However, regardless of the exact nature of the relationship between $I$ and $A_{V}$, an accurate measurement of the $p-I$ relationship is required in order to interpret dust grain alignment properties.

It is expected that observations of dense material within molecular clouds will show a power-law dependence of $p$ on $I$, $p\propto I^{-\alpha}$ \citep{whittet2008}.  This index is expected to steepen as grains become increasingly poorly aligned with the magnetic field: an index of $\alpha=0$ would indicate equal grain alignment at all optical depths, an index of $\alpha=0.5$ is predicted for a cloud in which grain alignment decreases linearly with increasing optical depth, while an index of $\alpha=1$ is predicted for an environment in which all observed polarized emission is produced in a thin layer at the surface of the cloud, and grains at higher densities have no preferred alignment relative to the magnetic field (c.f. \citealt{whittet2008}; \citealt{jones2015}).

Several recent studies of molecular clouds and starless cores have found power-law indices $\alpha$ in the range $0.5$ to $1$, for example: $\alpha=0.7$ in OMC-3 and $0.8$ in Barnard 1 \citep{matthews2000}; $0.64\pm0.01$ in CB 54 and $0.55\pm0.22$ in DC253-1.6 \citep{henning2001}; $0.83\pm0.01$ in NGC 2024 FIR 5 \citep{lai2002}; $0.92\pm 0.17$ in Pipe 109 \citep{alves2014,alves2015}; $0.84$ to $1.02$ in W51 \citep{koch2018}; 1.0 in FeSt 1-457 \citep{kandori2018}; $0.7$ or $0.8$ in Oph A \citep{kwon2018}; $0.9$ in Oph B \citep{soam2018}; and $1.0$ in Oph C \citep{liu2019}.  (Note that a larger value of $\alpha$ indicates a steeper negative slope.)  In most of these cases the data used have been selected to have signal-to-noise $\geq 3$ in some combination of polarization fraction, polarized intensity, and total intensity.  Recent improvements in instrumental sensitivity have allowed more stringent selection criteria: some recent observations have employed a signal-to-noise cut of 20 in total intensity \citep{kwon2018, soam2018}.  These results have generally been taken to suggest poor grain alignment within the densest parts of molecular clouds.

Polarized intensity and polarization fraction are both constrained to be positive quantities, and so  are characterized by Ricean statistics \citep{rice1945, serkowski1958}, albeit approximately so in the case of polarization fraction, as discussed below.  This results in a strong positive bias in measured polarization fraction at low total intensities.  Several methods of correcting for this bias have been proposed, of varying levels of sophistication \citep{wardle1974, simmons1985, vaillancourt2006, quinn2012, montier2015a, montier2015b, vidal2016, muller2017}.  These methods are collectively known as (statistical) debiasing.  An alternative approach is to bypass the problem of characterizing polarization fraction by working with the Stokes parameters of observed polarized emission directly (e.g. \citealt{herron2018}).  The Rice distribution is discussed extensively in the electrical engineering literature, due to its relevance to signal processing (e.g. \citealt{lindsey1964}; \citealt{sijbers1998}; \citealt{abdi2001}).

In this work we investigate the extent to which measurements of the $p-I$ index are biased by the statistical behavior of polarization fraction and by choice of selection criteria.  We demonstrate a method by which the relationship between $p$ and $I$ can be accurately characterized using the full observed data set, and without recourse to statistical debiasing.

The L1688 region of the Ophiuchus Molecular Cloud is a nearby ($138.4\pm 2.6$\,pc; \citealt{ortizleon2018}) site of low-to-intermediate-mass star formation \citep{wilking2008}.  The region contains a number of dense clumps, Oph A -- F, notable for their differing properties and star formation histories \citep{motte1998,pattle2015}.  The Oph A, B and C clumps have been observed with the James Clerk Maxwell Telescope (JCMT) POL-2 polarimeter as part of the JCMT BISTRO (B-Fields in Star-forming Region Observations) Survey (\citealt{wardthompson2017, kwon2018, soam2018, liu2019}).  We apply the methods developed in this paper to the JCMT BISTRO Survey observations of L1688.

This paper is structured as follows: in Section~\ref{sec:equations} we present the key equations governing the behavior of polarization fractions.  In Section~\ref{sec:model}, we present a simple model for the behavior of polarization fraction as a function of signal-to-noise.  In Section~\ref{sec:monte} we present Monte Carlo simulations demonstrating the behavior of our model and testing various fitting methods to recover the underlying relationship between polarization fraction and total intensity.  In Section~\ref{sec:data}, we apply our model to recent JCMT POL-2 observations of the Ophiuchus Molecular Cloud.  Section~\ref{sec:summary} summarizes our results.

\section{Mathematical properties of polarization fraction} \label{sec:equations}

Linearly polarized intensity is given by
\begin{equation}
P = \sqrt{Q^{2} + U^{2}},
\label{eq:pi}
\end{equation}
where $Q$ is the Stokes $Q$ intensity and $U$ is the Stokes $U$ intensity.  We do not here consider circular polarization, and so assume Stokes $V$ intensity to be zero throughout.  Polarization fraction, the ratio of polarized intensity $P$ to total intensity $I$, is given by
\begin{equation}
p = \frac{\sqrt{Q^{2} + U^{2}}}{I}.
\label{eq:polfrac}
\end{equation}

We take measurements of Stokes $I$, $Q$ and $U$ to have measurement errors of $\delta I$, $\delta Q$ and $\delta U$ respectively.  We assume that these measurement uncertainties are drawn from Gaussian distributions of width $\sigma_{I}$, $\sigma_{Q}$ and $\sigma_{U}$.

Note that all symbols defined and used in this work are summarized in Table~\ref{tab:symbols} in the Appendix.

\subsection{The Rice distribution}
\label{sec:equations_rice}

Polarized intensity -- the result of addition in quadrature of real numbers, as shown in equation~\ref{eq:pi} -- must be a positive quantity.  The addition in quadrature of small values of $Q$ and $U$, with measurement uncertainties $\delta Q$ and $\delta U$ will, where $\delta Q \gtrsim |Q|$ and $\delta U \gtrsim |U|$ produce spurious measurements of polarized intensity.  The results of such addition in quadrature are described mathematically by the Rice distribution \citep{rice1945}, where the quantities under addition have noise properties which are Gaussian and uncorrelated. In principle, we can expect uncertainties on Stokes $Q$ and $U$ to be independent, as $Q$ and $U$ are orthogonal components of the Stokes polarization vector (see, e.g., \citealt{herron2018}).  We discuss the validity of this assumption in the specific case of JCMT POL-2 observations in Section~\ref{sec:data}.

The Rice distribution has positive skewness at low signal-to-noise ratio (SNR), while at high SNR it tends towards a Gaussian distribution \citep{rice1945}.  Assuming that Stokes $Q$ and $U$ data have independent Gaussian measurement uncertainties, polarized intensities derived using equation~\ref{eq:pi} will be Rice-distributed (\citealt{wardle1974}; \citealt{simmons1985}).  

As the distributions of observed values of polarized intensity, and so of polarization fraction, are different from the underlying distributions which would be seen in the absence of measurement error, we henceforth denote observed values of polarized intensity and polarization fraction as $P^{\prime}$ and $p^{\prime}$, respectively.

The division $p^{\prime} = P^{\prime}/I$ (c.f. equation~\ref{eq:polfrac}) can, at low SNR (where $I$ is small and $P^{\prime}$ may be artificially large), produce artificially high values of $p^{\prime}$.  If $P^{\prime}/\delta P\ll I/\delta I$, as is generally the case as $p\ll 1$ in physically realistic situations, the effect of the measurement error on $I$ on the distribution of $p^{\prime}$ will be minimal, and so $p^{\prime}$ will also be approximately Rice-distributed (cf. \citealt{simmons1985}).  

If the polarization fractions which we measure can be treated as being Rice-distributed, the probability of measuring a polarization fraction $p^{\prime}$ given a true polarization fraction $p$ and an RMS uncertainty in polarization fraction $\sigma_{p}$ (the probability density function) is
\begin{equation}
PDF_{Rice}(p^{\prime}|p) = \frac{p^{\prime}}{\sigma_{p}^{2}}\exp\left(-\frac{p^{\prime 2}+p^2}{2\sigma_{p}^{2}}\right)\mathcal{I}_{0}\left(\frac{p^{\prime}p}{\sigma_{p}^{2}}\right),
\label{eq:pdf}
\end{equation}
where $\mathcal{I}_0$ is the zeroth-order modified Bessel function \citep{simmons1985}.  See \citet{montier2015a} for a derivation of this result (see also \citealt{bastien2007}; \citealt{hull2015}).

The mean of the Rice distribution is given by
\begin{equation}
\mu_p = \sqrt{\frac{\pi}{2}}\sigma_{p}\mathcal{L}_{\frac{1}{2}}\left(-\frac{p^{2}}{2\sigma_{p}^{2}}\right),
\label{eq:ricemean}
\end{equation}
where $\mathcal{L}_{\frac{1}{2}}$ is a Laguerre polynomial of order $\frac{1}{2}$:
\begin{multline}
\mathcal{L}_{\frac{1}{2}}\left(-\frac{p^{2}}{2\sigma_{p}^{2}}\right) = e^{-\frac{p^{2}}{4\sigma_{p}^{2}}}\left[\left(1+\frac{p^{2}}{2\sigma_{p}^{2}}\right)\,\mathcal{I}_{0}\left(\frac{p^{2}}{4\sigma_{p}^{2}}\right) \right. \\ \left. + \frac{p^{2}}{2\sigma_{p}^{2}}\,\mathcal{I}_{1}\left(\frac{p^{2}}{4\sigma_{p}^{2}}\right)\right],
\label{eq:laguerre}
\end{multline}
and $\mathcal{I}_{0}$ and $\mathcal{I}_{1}$ are modified Bessel functions of order 0 and 1, respectively.

We note that approximating $p^{\prime}$ as Rice-distributed is a statement that $I$ is perfectly known, and so equivalent to setting its RMS uncertainty $\sigma_{I}=0$.  We will demonstrate in Section~\ref{sec:monte} that equation~\ref{eq:pdf} describes synthetic data with realistic noise properties sufficiently well to justify making this assumption.

Having chosen to treat $I$ as perfectly known, if measurement uncertainties on $Q$ and $U$ are equivalent, i.e. $\sigma_{Q} = \sigma_{U}$, then $\sigma_{p}$ will be given by
\begin{equation}
    \sigma_{p} = \frac{\sigma_{Q}}{I} = \frac{\sigma_{U}}{I}.
    \label{eq:sigmap}
\end{equation}
(c.f. equation~\ref{eq:polfrac}).  This approximation is discussed by \citet{montier2015a}.

\subsection{Debiasing}
\label{sec:equations_debias}

Correction of observed polarization fraction $p^{\prime}$ for the bias described above is known as `debiasing'.  A commonly-used method for debiasing polarization fractions is the \citet{wardle1974} estimator (see also \citealt{serkowski1962}), under which debiased observed polarized intensity is given by
\begin{equation}
P^{\prime}_{db} =\sqrt{Q^{2} + U^{2}-\frac{1}{2}(\delta Q^{2} + \delta U^{2})}.
\end{equation}
Similarly, debiased observed polarization fraction is given by
\begin{equation}
p^{\prime}_{db} = \frac{\sqrt{Q^{2} + U^{2}-\frac{1}{2}(\delta Q^{2} + \delta U^{2})}}{I}.
\label{eq:polfrac_db}
\end{equation}

As this is the default debiasing method for POL-2 data (e.g. \citealt{kwon2018}), we compare the results of fitting non-debiased data and data debiased using the \citet{wardle1974} estimator in Section~\ref{sec:monte}.  For a detailed comparison of debiasing methods see \citet{montier2015b}.

\subsection{An aside on polarization angle}
\label{sec:equations_angle}

Polarization angle $\theta_{p}$ is given by
\begin{equation}
    \theta_{p} = \dfrac{1}{2}\arctan(U,Q).
\end{equation}
We note that measurements of polarization angle follow a different probability distribution than those of polarization fraction \citep{naghizadehkhouei1993}.  We expect measurements of polarization angle to be significantly more robust than those of polarization fraction, as the $Q$ and $U$ distributions have the same statistical properties and so their ratio is not affected by the issues discussed above.  Note also that polarization angle is not constrained to be positive.

Inference of grain alignment properties from polarization angle distribution would require at minimum a model of the plane-of-sky magnetic field morphology and an estimate of the fluctuations in magnetic field direction induced by Alfv\'{e}nic turbulence \citep{chandrasekhar1953a}.  We therefore do not consider polarization angles in this work.

\section{A simple model for polarization fraction} \label{sec:model}

\begin{figure*}
\centering
\includegraphics[width=0.7\textwidth]{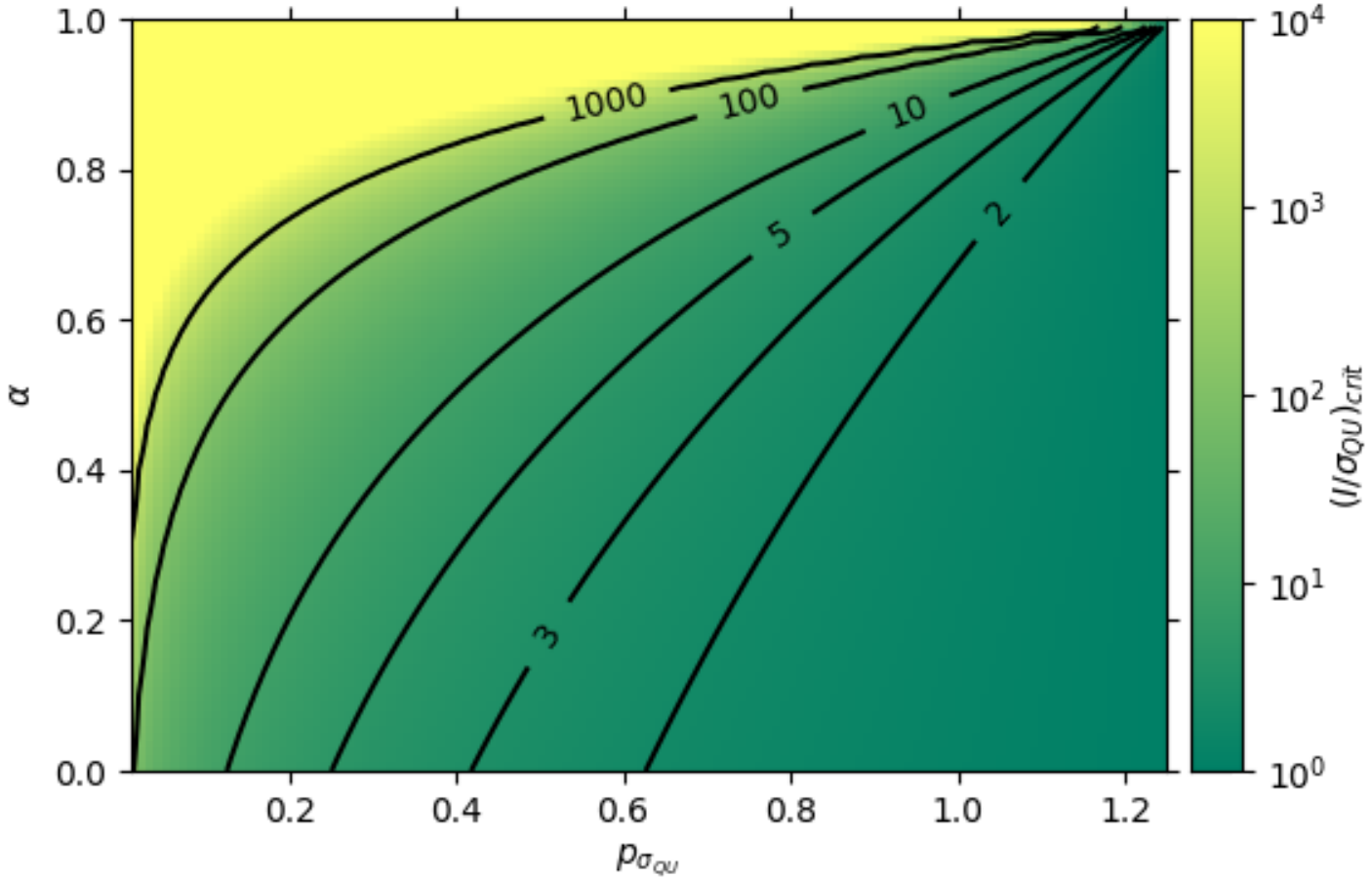}
\caption{The signal-to-noise value $(I/\sigma_{QU})_{crit}$ at which our model predicts a break in power-law index, as a function of $\alpha$ and $p_{\sigma_{QU}}$, derived using equation~\ref{eq:sn}.  In order to detect and accurately characterize a power-law behavior shallower than $\alpha=1$, a reasonable number of data points with $(I/\sigma_{QU})>(I/\sigma_{QU})_{crit}$ are required.  Contours show lines of constant SNR.  Color table saturates at $I/\sigma_{QU} = 10^{4}$.  Note that not all combinations of $\alpha$ and $p_{\sigma_{QU}}$ will produce physically plausible results.}
\label{fig:sn_limits}
\end{figure*}

We construct a simple model in which polarization fraction is fully described by
\begin{equation}
p(I) = p_{0}\left(\frac{I}{I_{0}}\right)^{-\alpha},
\label{eq:p_model}
\end{equation}
where $p_{0}$ is the polarization fraction at the reference intensity, $I_{0}$.  We expect $0 < \alpha \leq 1$ (e.g. \citealt{whittet2008}).

The reference intensity, or normalization, $I_{0}$, can be treated as a free parameter of the model. However, we choose to specify $I_{0} = \sigma_{QU}$, where $\sigma_{QU}$ is a single value, with units of intensity, representative of the RMS noise in both Stokes $Q$ and Stokes $U$ measurements.  We thus assume that the Stokes $Q$ and $U$ data sets have identical statistical properties, and can be adequately characterized by a single RMS noise value.  This is a reasonable assumption for recent submillimeter emission polarization measurements, as described in Section~\ref{sec:data}, below.

By taking $I_{0} = \sigma_{QU}$, we assume that the power-law relationship between $I$ and $p$ applies to all measurements above the noise level of the data.  Choosing $I_{0} = \sigma_{QU}$ allows us to discuss the behavior of our model in terms of a simple signal-to-noise criterion, $I/\sigma_{QU}$, as demonstrated below.  The reference polarization fraction $p_{0}$ thus depends on $\sigma_{QU}$.  In order to emphasize this, we define the polarization fraction at the noise level of the data, $p(I = \sigma_{QU}) = p_{\sigma_{QU}}$, i.e. if $I_{0}=\sigma_{QU}$, $p_{0}=p_{\sigma_{QU}}$.

In order to meaningfully compare polarization fractions in data sets with different RMS noise levels, we must convert $p_{\sigma_{QU}}$ to polarization fraction at a common reference intensity level.  In this work, we choose a reference intensity of $I=100$\,mJy/beam, and so define
\begin{equation}
p_{100\,{\rm mJy/beam}} = p_{\sigma_{QU}}\left(\dfrac{100\,{\rm mJy/beam}}{\sigma_{QU}}\right)^{-\alpha}.
\label{eq:p100}
\end{equation}
The choice to reference to 100\,mJy/beam is largely arbitrary, but suitable for the JCMT POL-2 data which we consider in Section~\ref{sec:data} below, in which $I=100$\,mJy/beam is both significantly above the RMS noise level of the data, and below the maximum intensities observed.  Note that the JCMT has an effective beam size of 14.1 arcsec at 850\,$\mu$m \citep{dempsey2013}.

Equation~\ref{eq:p_model} becomes unphysical where $p(I)>1$.  Moreover, in physically realistic scenarios we expect $p\lesssim 0.2$ in data with good signal-to-noise, as the maximum percentage polarization observed in the diffuse ISM is $\sim 20$\% \citep{planck2015}.  We thus require that $p < 1$, and expect that $p\lesssim0.2$, wherever we believe our observed Stokes $Q$, $U$ and $I$ values to be reliable.  

\subsection{Physical implications of $\alpha$}
\label{sec:model_alpha}

If $\alpha = 0$, this implies that polarized intensity $P = \sqrt{Q^{2} + U^{2}} \propto I$, and so that $p = p_{\sigma_{QU}}$ everywhere.  If polarized intensity is directly proportional to total intensity, this indicates that all emission along each sightline is polarized to the same degree, and so that there is no variation in polarization efficiency anywhere within the observed cloud.

If $\alpha = 1$, this implies that $P$ does not vary with $I$; i.e. a constant amount of polarized emission is observed at all locations in the cloud.  This indicates that only a small portion of the total line of sight is contributing polarized emission.  This is usually interpreted as a thin layer of polarized emission overlaying an otherwise unpolarized optically thin sightline.  This polarized emission is implicitly from low-density material, and so, if the higher-density material along the sightline were also polarized, that polarized emission ought also to be detectable.  Thus, claiming a $physical$ index of $\alpha=1$ is essentially a statement that one has observed statistical noise in Stokes $Q$ and $U$ around a constant level of polarized intensity, and that this indicates a genuine absence of polarized emission at high total intensities, rather than insufficient SNR to achieve a detection.

Intermediate values of $0<\alpha<1$ indicate a positive relationship, $P\propto I^{1-\alpha}$, in which the total amount of polarized emission increases with, but slower than, total emission.  This implies that as the amount of material along the sightline increases, the ability of that material to produce polarized emission decreases (under the assumptions of isothermal, optically thin emission).  Thus, an index $0<\alpha<1$ indicates that depolarization increases with depth into the cloud.
 
\subsection{Low-SNR limit}
\label{sec:model_lowsn}

In the low-SNR limit, the distribution of $p^{\prime}$ tends to the Rice distribution regardless of the value of $\alpha$ (or of $p_{0}$ or $I_{0}$).  In the limit where $p << \sigma_p$, $\mathcal{L}_{\frac{1}{2}}(-p^{2}/2\sigma_p)\to 1$, and therefore $\mu_{p}\to \sigma_{p}\sqrt{\pi/2}$.  Taking $\sigma_{p} = \sigma_{QU}/I$, the behavior of the distribution is well approximated by
\begin{equation}
p^{\prime} = \sqrt{\frac{\pi}{2}}\left(\frac{I}{\sigma_{QU}}\right)^{-1}
\label{eq:lowlim}
\end{equation}
at small $I$. Thus, at low SNR, $p^{\prime}\propto I^{-1}$, regardless of the true value of $\alpha$.

In the no-signal case ($p_{\sigma_{QU}} = 0$; $Q=U=0$), equation~\ref{eq:lowlim} characterizes the observed signal at all values of $I$.  \emph{Thus, in the absence of a true measurement, a $p^{\prime} \propto I^{-1}$ behavior will be observed.}

\subsection{High-SNR limit}
\label{sec:model_highsn}

In the high-SNR limit, the distribution of $p^{\prime}$ tends toward a Gaussian distribution around the true value,
\begin{equation}
p^{\prime} = p_{\sigma_{QU}}\left(\frac{I}{\sigma_{QU}}\right)^{-\alpha}.
\label{eq:highlim}   
\end{equation}	

\subsection{SNR criterion}
\label{sec:model_SNcrit}

Equations~\ref{eq:lowlim} and \ref{eq:highlim} suggest that, for $\alpha < 1$, the underlying power-law dependence of $p$ on $I$ will be observable when
\begin{equation}
\frac{I}{\sigma_{QU}} > \left(\frac{I}{\sigma_{QU}}\right)_{crit} = \left(\frac{1}{p_{\sigma_{QU}}}\sqrt{\frac{\pi}{2}}\right)^{\frac{1}{1-\alpha}}.
\label{eq:sn}
\end{equation}
At SNRs below this critical value, the artificial $p^{\prime}\propto I^{-1}$ behaviour will dominate, and the true value of $\alpha$ will not be recoverable.

Figure~\ref{fig:sn_limits} shows solutions of equation~\ref{eq:sn} for the critical SNR value $(I/\sigma_{QU})_{crit}$, above which the true power-law behavior (i.e. the true value of $\alpha$) would be recoverable.  It is apparent that significant SNR is required to distinguish a power-law behavior with a steep value of $\alpha$ from instrumental noise.  

\begin{figure}
\centering
\includegraphics[width=0.47\textwidth]{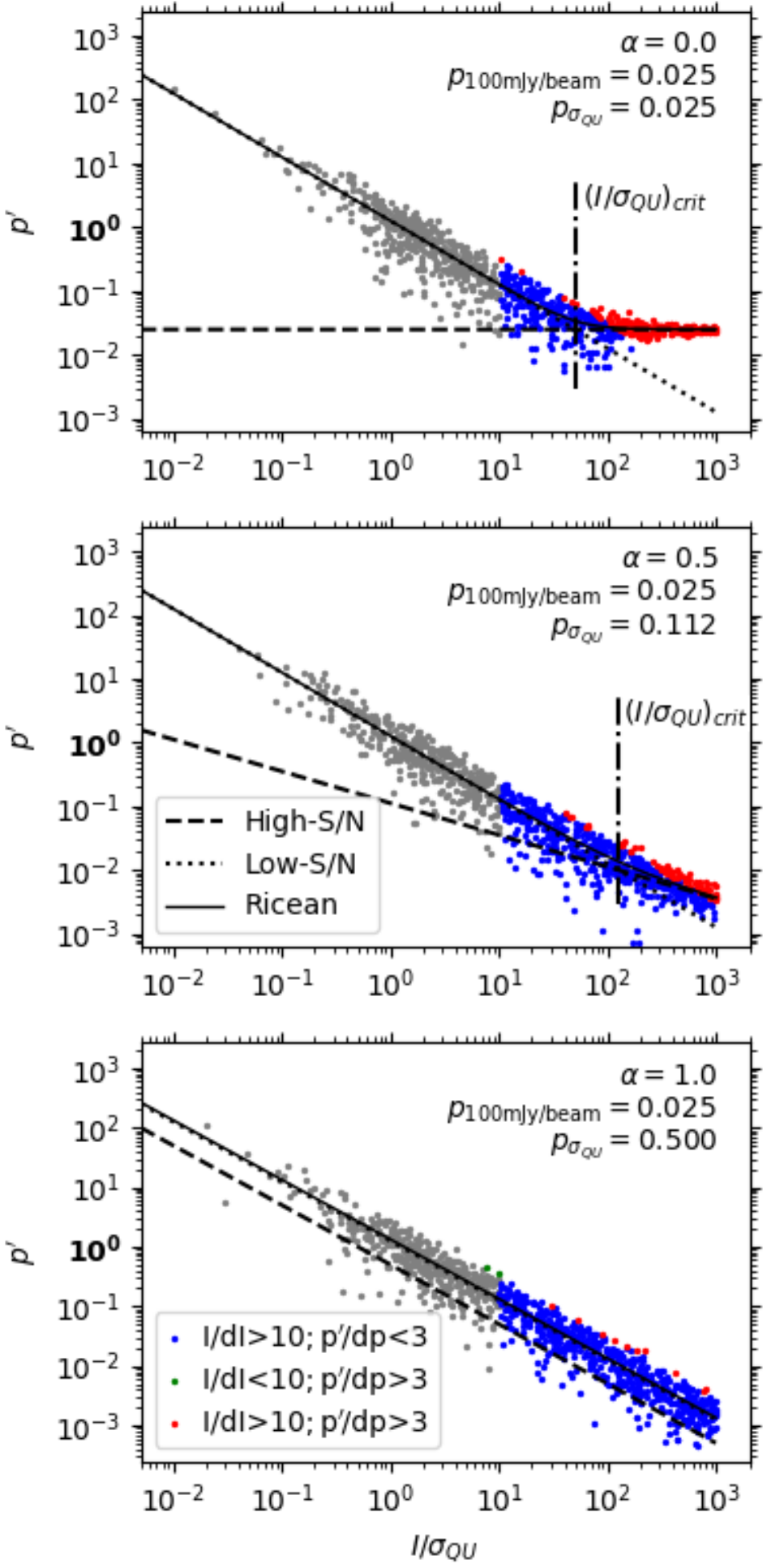}
\caption{Monte Carlo simulations of the observed polarization fraction $p^{\prime}$ as a function of signal-to-noise $I/\sigma_{QU}$ for three cases: $\alpha = 0$ (top), $\alpha=0.5$ (middle), and $\alpha = 1$ (bottom).  In all cases, $p_{100{\rm mJy/beam}}=0.025$.  Results are shown without debiasing.  Polarization fraction $p^{\prime}$ is given as an absolute value (not as a percentage), and so any value $p^{\prime}>1$ (marked in bold on plot axes) is unphysical.  Points are colored according to their signal-to-noise in $I$ and $p^{\prime}$; grey points meet none of the criteria shown in the legend.  The solid black line shows $p^{\prime}=\mu_{p}$, the mean of the Rice distribution, for the given values of $\alpha$ and $p_{\sigma_{QU}}$.  The dashed black line shows the true polarization fraction $p = p_{\sigma_{QU}}(I/\sigma_{QU})^{-\alpha}$.  The dotted line shows the null-hypothesis/low-SNR relation $p^{\prime} = \sqrt{\pi/2}(I/\sigma_{QU})^{-1}$.  In the cases where $\alpha<1$, the dot-dashed line marks the critical value $(I/\sigma_{QU})_{crit}=(p_{\sigma_{QU}}^{-1}\sqrt{\pi/2})^{\frac{1}{1-\alpha}}$.}
\label{fig:monte1}
\end{figure}

\section{Monte Carlo Simulations} \label{sec:monte}

In order to test the accuracy of our interpretation of the model described above, we performed a set of Monte Carlo simulations.

We investigated values of $\alpha$ in the range $0\leq\alpha\leq 1$.  In each case we took $p_{100{\rm mJy/beam}} = 0.025$ (i.e. emission is intrinsically 2.5\% polarized at $I=100.0$ -- intensity units are arbitrary, but chosen to mimic POL-2 data, which are calibrated in mJy/beam).  This value of $p_{100{\rm mJy/beam}}$ was chosen for similarity to POL-2 observations.  We then calculated the appropriate value of $p_{\sigma_{QU}}$ for our chosen $\alpha$ and $p_{100{\rm mJy/beam}}$ using equation~\ref{eq:p100}.

As we are not concerned with polarization angles in this work, we choose for simplicity that Stokes $U = 0$ ($\theta_{p}=0^{\circ}$, implying that the underlying magnetic field is uniform, and oriented $90^{\circ}$ east of north), and so equation~\ref{eq:p_model} is equivalent to 
\begin{equation}
Q(I) = p(I)\times I = p_{\sigma_{QU}}\sigma_{QU}^{\alpha}I^{1-\alpha}.
\end{equation}

We drew a set of $10^{3}$ randomly distributed $\log_{10}(I/\sigma_{QU})$ values in the range $0 < \log_{10}(I/\sigma_{QU}) < 3$ (thus assuming that low-SNR values of $I$ are more probable).  We drew measurement errors $\delta Q$ and $\delta U$ on Stokes $Q$ and $U$ from Gaussian distributions of equal (fixed) width, $\sigma_{QU}$, and drew measurement errors $\delta I$ on Stokes $I$ from a Gaussian distribution of (fixed) width $\sigma_{I}$. We here chose $\sigma_{QU} = \sigma_{I} = 5.0$ (arbitrary units).  Equations~\ref{eq:pdf} and \ref{eq:ricemean} assume that the effect of $\sigma_{I}$ on the distribution of $p$ is negligible.  We wished to test whether that approximation is valid in observations where noise on Stokes $I$ is comparable to that on Stokes $Q$ and $U$.

The `observed' values $Q_{obs}$, $U_{obs}$ and $I_{obs}$ were then, for the $i^{{\rm th}}$ value  
\begin{eqnarray}
Q_{obs,i} \,& \,=\,  & \,Q_{i}  + \, \delta Q_{i} \\
U_{obs,i} \,& \,= \, & \,\delta U_{i} \\
I_{obs,i} \,& = & \, I_{i} + \delta I_{i}.
\end{eqnarray}
`Observed' values of $p^{\prime}$ and $p^{\prime}_{db}$ were calculated using equations \ref{eq:polfrac} and \ref{eq:polfrac_db} respectively, for a given value of $p_{\sigma_{QU}}$. We repeated this process $10^{4}$ times.

Figure~\ref{fig:monte1} shows a single realization of our Monte Carlo simulations for three cases:  $\alpha = 1$, $\alpha = 0.5$, and $\alpha=0$ (constant polarization fraction).  In all cases, $p_{100{\rm mJy/beam}} = 0.025$.  Results are shown without any debiasing of polarization fraction, and are colored according to their signal-to-noise in $I$ and $p^{\prime}$.

Figure~\ref{fig:monte1} shows that where $I/\sigma_{QU} <(I/\sigma_{QU})_{crit}$, the distribution shows an identical $p^{\prime}\propto I^{-1}$ behavior in all cases, and that the underlying power-law behavior does not dominate over the $I^{-1}$ dependence until $I/\sigma_{QU} \gg (I/\sigma_{QU})_{crit}$.  Thus unless a reasonable number of data points have signal-to-noise significantly greater than the critical value, the true power-law behavior is not recoverable.  The overall behavior of the recovered polarization fraction is well-described by the mean of the Rice distribution.

\begin{figure*}
\centering
\includegraphics[width=\textwidth]{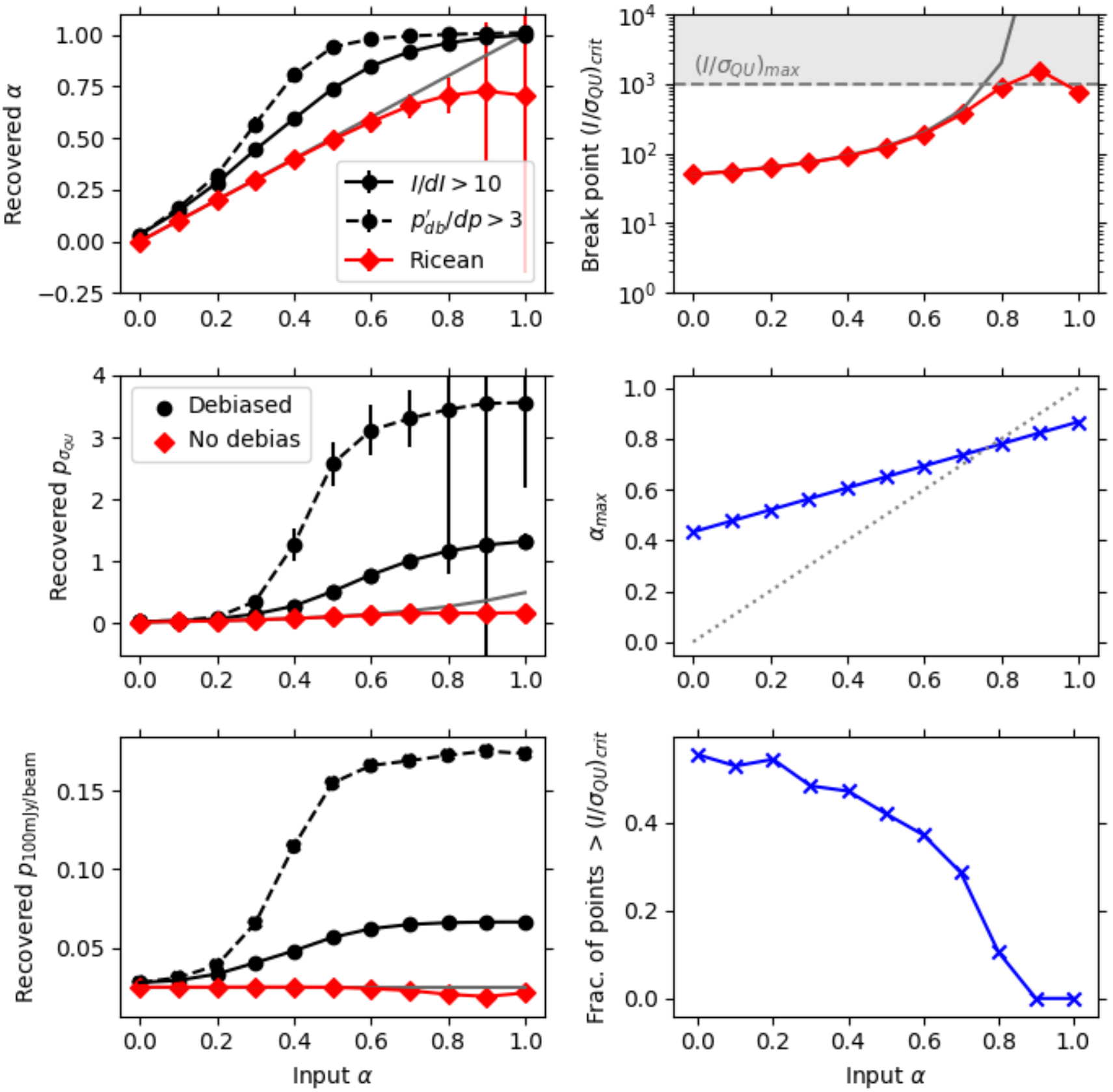}
\caption{Results of fitting to Monte Carlo simulations, for $p_{100{\rm mJy/beam}} = 0.025$ and $0\leq \alpha \leq 1$.  \textit{Left-hand column:} Black data points show single-power-law fitting to debiased data where $p/\delta p>3$ (dashed line) and $I/\delta I>10$ (solid line).  Red data points show fitting of the Ricean-mean model to non-debiased data.  Top row: recovered values of $\alpha$; middle row: recovered values of $p_{\sigma_{QU}}$; bottom row: recovered values of $p_{100{\rm mJy/beam}}$.  All rows: Solid grey line shows input values.  \textit{Right-hand column (relevant to Ricean-mean model only)}: Top row: critical SNR value $(I/\sigma_{QU})_{crit}$ inferred from best-fit $\alpha$ and $p_{\sigma_{QU}}$, for the Ricean-mean model.  Solid grey line shows input values.  Horizontal dashed line shows the maximum signal-to-noise value in the simulations, $(I/\sigma_{QU})_{max} = 1000$, and values $(I/\sigma_{QU})_{crit} > (I/\sigma_{QU})_{max}$ (indicating the range in which input model values are not recoverable) are shaded in grey.  Middle row: maximum recoverable index $\alpha_{max}$.  Dotted grey line shows the 1:1 relation; values near or to the right of this line will not be accurately recoverable.  Bottom row: Fraction of data points above the critical SNR $(I/\sigma_{QU})_{crit}$.}
\label{fig:fitting}
\end{figure*}

\subsection{Uncertainty on polarization fraction}
\label{sec:empiric_uncert}

In keeping with standard practice in observational polarimetry, and the default behavior of the POL-2 pipeline (e.g. \citealt{kwon2018}), we estimated uncertainty on polarization fraction using the relation
\begin{equation}
\delta p = \left(\frac{Q^{2}\delta Q^{2} + U^{2}\delta U^{2}}{I^{2}(Q^{2}+U^{2})} + \frac{\delta I^{2}(Q^{2}+U^{2})}{I^{4}}\right)^{\frac{1}{2}},
\label{eq:dp}
\end{equation} 
(e.g. \citealt{wardle1974}).  We note that this results is derived using classical error propagation and so assumes that $\delta Q$, $\delta U$ and $\delta I$ are small and uncorrelated.  As discussed above, measurement errors $\delta p$ calculated using equation~\ref{eq:dp} can only be treated as representative of a Gaussian distribution around $p^{\prime}$ when $p^{\prime}/\delta p$ is large.

\subsection{Fitting methods}
\label{sec:monte_fit}

We investigated the results of fitting two models to our Monte Carlo simulations.  Fitting was performed using the \emph{scipy} routine \emph{curve\_fit}.  In all cases we attempted to recover the input values of $p_{\sigma_{QU}}$ and $\alpha$, assuming that $\sigma_{QU}$ is a fixed and directly measurable property of the data set.  We supplied our calculated $\delta p$ values to \textit{curve\_fit} as 1-$\sigma$ uncertainties on $p^{\prime}$.  Although this is technically valid only at high SNR, the default behavior of this fitting routine is to consider only the relative magnitudes of the input uncertainties.  The uncertainties provided thus down-weight the contribution of low-SNR points to the fitting process.

\paragraph{Single power-law}  In keeping with standard practice, we fitted a single power-law model,
equation~\ref{eq:highlim}, to the high-SNR data.  This model assumes that $p^{\prime}\approx p$, and so is usually applied to data which has been statistically debiased.  Applying this model to data which had not been debiased would inherently lead to artificially high values of $\alpha$ being recovered.  In order to fairly test this model, we thus applied it to the debiased polarization fractions $p^{\prime}_{db}$ returned by our Monte Carlo simulations, i.e. we fitted
\begin{equation}
p^{\prime}_{db} = p_{\sigma_{QU}}\left(\frac{I}{\sigma_{QU}}\right)^{-\alpha}
\label{eq:splmodel}
\end{equation}
to a high-SNR subset of the debiased data.  We tested two commonly-used SNR criteria: $I/\delta I> 10$ and $p^{\prime}_{db}/\delta p> 3$.  As we are here selecting higher-SNR data points, the $\delta p$ values which we use should be somewhat representative of 1-$\sigma$ Gaussian uncertainties on $p^{\prime}_{db}$.  However, this model will provide accurate values of $p_{\sigma_{QU}}$ and $\alpha$ only if the $p^{\prime}_{db}$ values selected are well within the high-SNR limit described in Section~\ref{sec:model_highsn}.

\paragraph{Mean of Rice distribution}  We fitted the mean of the Rice distribution (equation~\ref{eq:ricemean}), assuming that $p$ is given by equation~\ref{eq:highlim}, and that $\sigma_{p}\approx\sigma_{QU}/I$, i.e.:
\begin{equation}
p^{\prime} = \sqrt{\frac{\pi}{2}}\left(\frac{I}{\sigma_{QU}}\right)^{-1}\mathcal{L}_{\frac{1}{2}}\left[-\frac{p_{\sigma_{QU}}^{2}}{2}\left(\frac{I}{\sigma_{QU}}\right)^{2(1-\alpha)}\right].
\label{eq:ricemodel}
\end{equation}
We hereafter refer to this model as the `Ricean-mean model'.  This model is applied to the entire data set, without any selection by SNR, and is applied to data which has not been statistically debiased.  The Ricean-mean model is predicated on the Ricean probability density function (equation~\ref{eq:pdf}) being applicable to the data, which will not be the case if any attempt has been made to correct the data for observational bias.  We note that although the $\delta p$ values used in the fitting process do not strictly represent 1-$\sigma$ Gaussian uncertainties, they do have the effect of significantly down-weighting the contribution of low-SNR data to the best-fit model.  Fitting this model thus requires a good measurement of $\sigma_{QU}$ with which to constrain the low-SNR behavior.

We fitted these models to non-debiased (Ricean-mean) and debiased (power-law) values of $p^{\prime}$ for each realization of our Monte Carlo simulations.  We tested values of $\alpha$ in the range $0 \leq \alpha \leq 1$.  Our results are shown in Figure~\ref{fig:fitting}.  We find that of the fitting models which we tested, only the Ricean-mean model can accurately recover both $\alpha$ and $p_{\sigma_{QU}}$ when $\alpha$ is large.    

For data with a maximum $I/\sigma_{QU}$ value of 1000 and $p_{100{\rm mJy/beam}}=0.025$, $\alpha$ and $p_{\sigma_{QU}}$ can be approximately recovered with the single-power-law model only while $\alpha < 0.3$.  For the single-power-law model, the recovered value of $\alpha$ is systematically larger than the input value, tending towards 1 as $\alpha$ increases (thus, an input value $\alpha=1$ can be accurately recovered).  The recovered value of $p_{\sigma_{QU}}$ is also larger than the input value for steep values of $\alpha$.  Selecting $p/\delta p>3$ systematically increases the recovered value of $p_{\sigma_{QU}}$, as would be expected from examination of Figure~\ref{fig:monte1}.  Thus we expect fitting a single-power-law model to, in general, systematically return larger-than-input values of $\alpha$ and $p_{\sigma_{QU}}$, and so to overestimate the extent to which depolarization is occurring.  This is shown in Figure~\ref{fig:fitting}.

We find that the Ricean-mean model performs well for most values of $\alpha$.  As expected, the Ricean-mean model cannot accurately recover the input model parameters without a significant number of data points above $(I/\sigma_{QU})_{crit}$.  If the data set under consideration has a maximum SNR $(I/\sigma_{QU})_{max}$, a necessary condition for the input model parameters to be recoverable will be
\begin{equation}
    \left(\dfrac{I}{\sigma_{QU}}\right)_{max} > \left(\dfrac{I}{\sigma_{QU}}\right)_{crit},
    \label{eq:snmax}
\end{equation}
as if the break from $p^{\prime}\propto I^{-1}$ to $p^{\prime}\propto I^{-\alpha}$ occurs above the maximum SNR in the data set, $\alpha$ will perforce not be recoverable.  Thus, for a given value of $p_{\sigma_{QU}}$, there will be some theoretical maximum (i.e. steepest) recoverable value of $\alpha$, which we define as $\alpha_{max}$, associated with $(I/\sigma_{QU})_{max} = (I/\sigma_{QU})_{crit}$.  Combining this equality with equation~\ref{eq:sn} (substituting $\alpha_{max}$ for $\alpha$ in the latter), we find
\begin{equation}
\alpha_{max} = 1 - \frac{\log\left(\frac{1}{p_{\sigma_{QU}}}\sqrt{\frac{\pi}{2}}\right)}{\log\left(\left[\frac{I}{\sigma_{QU}}\right]_{max}\right)}.
\end{equation}
In practice, a reasonable number of data points must have SNRs greater than $(I/\sigma_{QU})_{crit}$ in order to accurately recover the input model values.  Figure~\ref{fig:fitting} shows that, for our chosen values of $p_{100{\rm mJy/beam}} = 0.025$, $\sigma_{QU} = 5.0$, and $(I/\sigma_{QU})_{max} = 1000$, the Ricean-mean model can accurately recover values up to $\alpha\approx 0.6$, and can approximately recover $\alpha\approx 0.7$, but that values steeper than this cannot be recovered accurately.  The $\alpha=0.9$ and $1.0$ cases have no data points above $(I/\sigma_{QU})_{crit}$.  Our results suggest that in order to accurately recover the input parameters, $\gtrsim 40$\% of the data points must be above $(I/\sigma_{QU})_{crit}$.

When performing these Monte Carlo simulations, we set $\sigma_{I}=\sigma_{QU}$.  We find that the Ricean-mean model can accurately recover the input model values despite the approximation $\sigma_{I}=0$ in equation~\ref{eq:pdf}.   This supports our assumption that the effect of uncertainty on total intensity on observed polarization fraction is negligible compared to that on polarized intensity.

We find that the Ricean-mean model consistently performs better than single-power-law fitting in accurately recovering $\alpha$ and $p_{\sigma_{QU}}$.  We therefore choose to apply this model to our observational data.

\begin{figure*}
\centering
\includegraphics[width=0.7\textwidth]{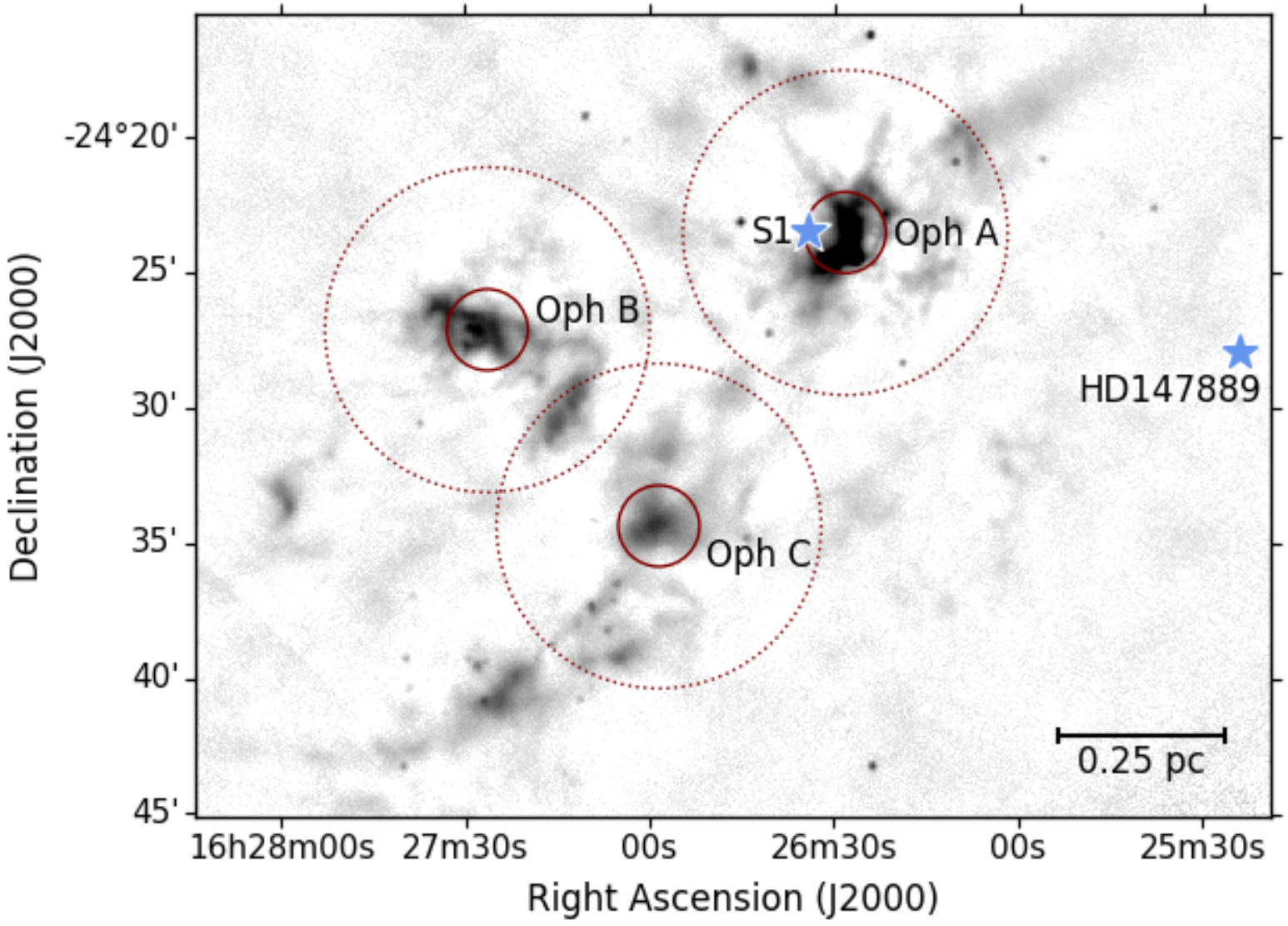}
\caption{A finding chart for the L1688 region.  The greyscale image shows the JCMT Gould Belt Survey SCUBA-2 850\,$\mu$m map of the region \citep{pattle2015}.  The three BISTRO fields are marked in red: solid lines show the central 3-arcminute-diameter regions with uniform noise characteristics, used in this work; dotted lines show the full extent of the observations.  The locations of the B stars HD147889 and S1 are marked with blue stars.  A plane-of-sky distance of 0.25\,pc at our assumed distance to L1688 of 138\,pc is shown in the lower right-hand corner.}
\label{fig:l1688_all}
\end{figure*}

\begin{figure*}
\centering
\includegraphics[width=\textwidth]{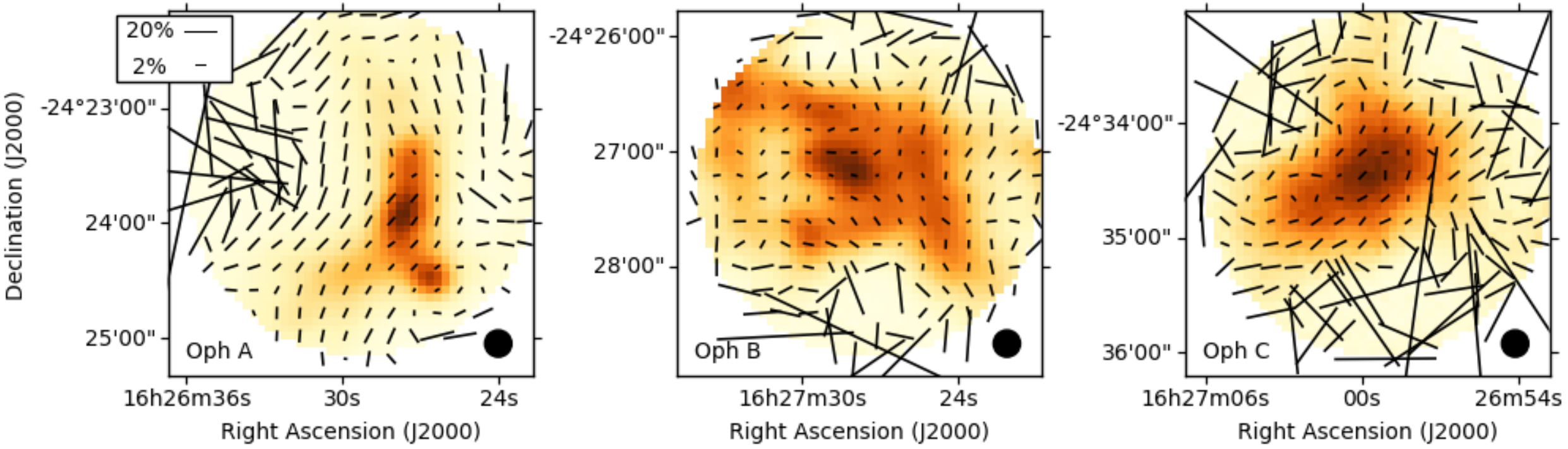}
\caption{POL-2 850$\mu$m total intensity maps of the central 3 arcminutes of the BISTRO Oph A, B and C fields, with polarization vectors overlaid.  Vectors are sampled to a 12-arcsecond grid and are scaled according to $\sqrt{p^{\prime}}$.  Polarization fractions have not been debiased.  20\% and 2\% polarization vectors are shown in the upper right-hand corner of the left-hand panel, demonstrating the non-linearity of the vector scale.  The JCMT 850$\mu$m beam size is shown in the lower right-hand corner of each plot.}
\label{fig:oph}
\end{figure*}

\section{An example: JCMT POL-2 observations of the L1688 region of the Ophiuchus Molecular Cloud} \label{sec:data}

\begin{table*}
\centering
\begin{tabular}{@{\extracolsep{2pt}}cccccccc@{}}
\hline
\hline
Pixel &  &  & Null & \multicolumn{4}{c}{Ricean-mean model} \\ \cline{4-4} \cline{5-8}
Size & $\langle\sigma_{QU}\rangle$ & $N$ & \multirow{2}{*}{$\displaystyle \frac{\chi^{2}}{N}$} &  & & & \multirow{2}{*}{$\displaystyle \frac{\chi^{2}}{N-2}$} \\
(arcsec) & (mJy/beam) & & & $\alpha$ & $p_{\sigma_{QU}}$ & $p_{100{\rm mJy/beam}}$ & \\
\hline
\hline
\multicolumn{8}{c}{Oph A} \\
\hline
4 & 3.29$\pm$0.55	& 1456	& 33.3	&  0.34$\pm$0.02 & 0.15$\pm$0.02	&  0.047$\pm$0.009	& 9.2	\\
8 & 1.88$\pm$0.36	& 400	& 101.0	& 0.34$\pm$0.03 & 0.18$\pm$0.04	& 0.047$\pm$0.016 & 24.9	\\
12	& 1.45$\pm$0.50	& 192	& 189.4	&  0.33$\pm$0.05 & 0.19$\pm$0.06	& 0.047$\pm$0.025 & 45.1	\\
16	& 1.15$\pm$0.44	& 114	& 291.9	&  0.34$\pm$0.06 & 0.21$\pm$0.09	& 0.046$\pm$0.032	& 73.0	\\
20	& 1.07$\pm$0.66	& 79	& 397.2	& 0.39$\pm$0.08 & 0.31$\pm$0.18	& 0.053$\pm$0.050	& 110.6	\\
24	& 0.95$\pm$0.66	& 56	& 437.1	& 0.38$\pm$0.10	&  0.30$\pm$0.20 & 0.051$\pm$0.058	& 98.9	\\
28	& 1.02$\pm$0.77	& 47	& 480.4	& 0.31$\pm$0.12	& 0.18$\pm$0.16 & 0.043$\pm$0.062	&  115.5	\\
32	& 0.60$\pm$0.16	& 31	& 832.0	& 0.00$\pm$0.16 & 0.02$\pm$0.03	& 0.020$\pm$0.046 & 172.1 \\
\hline
\multicolumn{8}{c}{Oph B} \\
\hline
4	&	3.37$\pm$0.57	&	1404 &	1.4	&	0.86$\pm$0.03	& 0.89$\pm$0.11	& 0.048$\pm$0.011	&	1.0	\\
8	&	1.93$\pm$0.41	&	390	&	2.5	&	0.78$\pm$0.05	& 0.73$\pm$0.18	& 0.034$\pm$0.015  &	1.4	\\
12	&	1.49$\pm$0.54	&	191	&	3.7	&	0.76$\pm$0.07	& 0.68$\pm$0.25	& 0.028$\pm$0.018  &	2.0	\\
16	&	1.18$\pm$0.49	&	115	&	4.7	&	0.70$\pm$0.09	& 0.53$\pm$0.26	& 0.024$\pm$0.021  &	2.2	\\
20	&	1.12$\pm$0.76	&	78	&	6.6	&	0.66$\pm$0.12	& 0.44$\pm$0.29	& 0.023$\pm$0.027 & 3.2	\\
24	&	0.99$\pm$0.73	&	58	&	7.7	&	0.71$\pm$0.12	& 0.58$\pm$0.38	& 0.022$\pm$0.026  &	3.5	\\
28	&	1.03$\pm$0.77	&	47	&	6.4	&	0.57$\pm$0.18	& 0.22$\pm$0.22	& 0.016$\pm$0.030  & 3.2	\\
32	&	0.61$\pm$0.16	&	32	&	12.2 &  0.59$\pm$0.16	& 0.34$\pm$0.34	& 0.017$\pm$0.030	&	4.2	\\
\hline
\multicolumn{8}{c}{Oph C} \\
\hline
4	&	3.70$\pm$0.65	&	1239	&	0.9	&	0.83$\pm$0.03	& 0.75$\pm$0.09	& 0.049$\pm$0.011	&	0.7	\\
8	&	2.12$\pm$0.45	&	359		&	1.4	&	0.84$\pm$0.04	& 0.93$\pm$0.16	& 0.037$\pm$0.012  &	1.0	\\
12	&	1.64$\pm$0.63	&	177		&	2.1	&	0.76$\pm$0.07	& 0.70$\pm$0.20	& 0.031$\pm$0.018	&	1.3	\\
16	&	1.30$\pm$0.54	&	109	    &	2.3	&	0.75$\pm$0.07	& 0.72$\pm$0.24	& 0.027$\pm$0.018	&	1.3	\\
20	&	1.22$\pm$0.80	&	75      &	2.6	&	0.69$\pm$0.09	& 0.52$\pm$0.22	& 0.025$\pm$0.020	&	1.3	\\
24	&	1.10$\pm$0.87	&	55	    &	2.9	&	0.71$\pm$0.12	& 0.52$\pm$0.29	& 0.021$\pm$0.023	&	1.7	\\
28	&	1.15$\pm$0.88	&	45  	&	4.0	&	0.58$\pm$0.14	& 0.30$\pm$0.21	& 0.023$\pm$0.030	&	2.0	\\
32	&	0.67$\pm$0.18	&	32	    &	6.0	&	0.63$\pm$0.11	& 0.52$\pm$0.30	& 0.022$\pm$0.025	&	2.0	\\
\hline
\hline
\end{tabular}
\caption{Mean RMS noise in Stokes $Q$ and $U$, $\langle\sigma_{QU}\rangle$, number of pixels, $N$, and fitting results for the Ricean-mean model, for L1688 Oph A, B and C data on pixel sizes of 4 to 32 arcsec.  Note that results for Oph A on 32-arcsec pixels are not reliable.}
\label{tab:l1688}
\end{table*}

\begin{figure}
\centering
\includegraphics[width=0.47\textwidth]{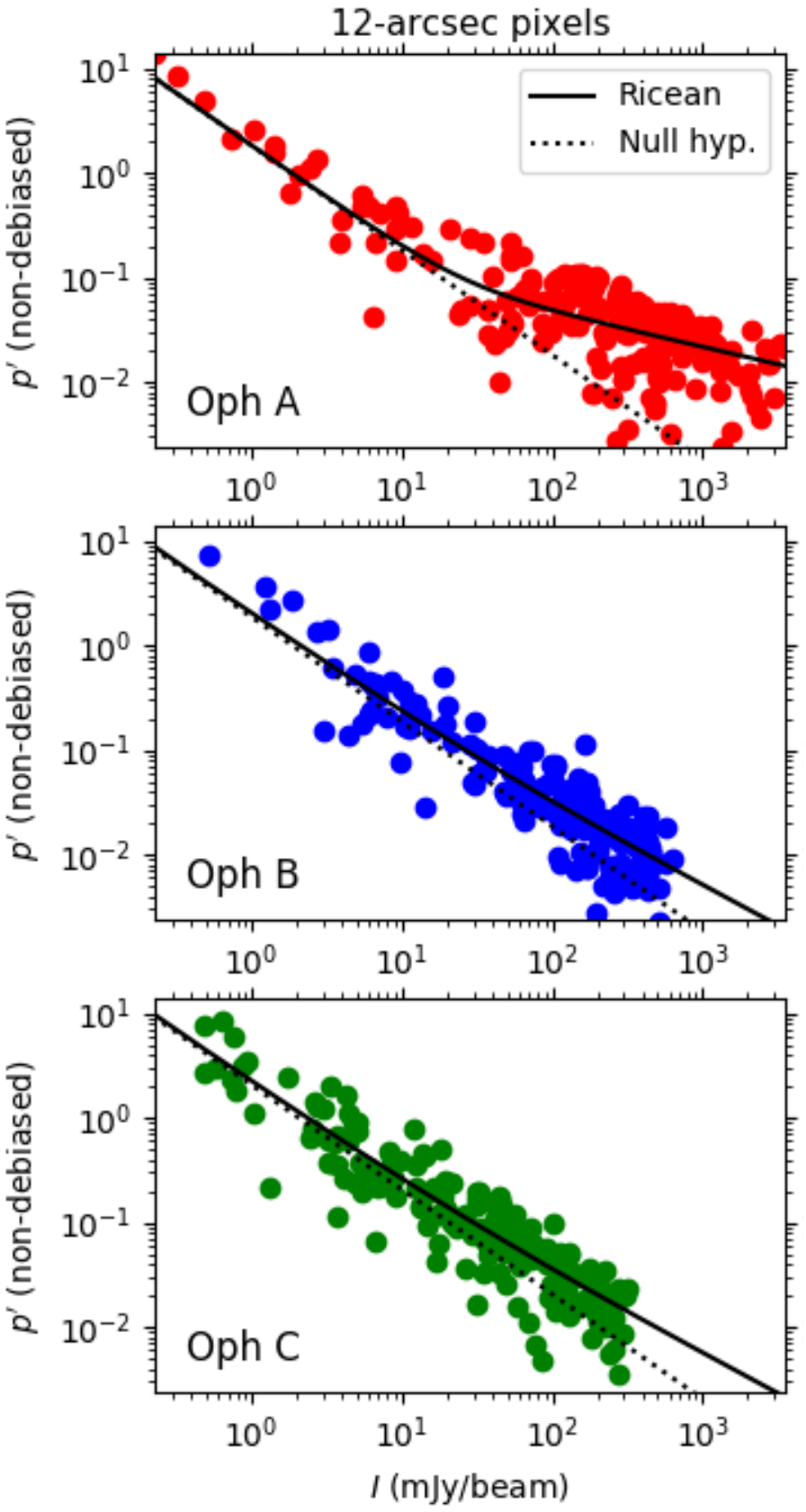}
\caption{A plot of observed polarization fraction $p^{\prime}$ versus total intensity $I$ for the Oph A, B and C data, gridded onto 12-arcsec pixels.  Top row: Oph A, middle row: Oph B, bottom row: Oph C.  Solid black line shows the best-fitting Ricean-mean model.  Dotted black line shows the expected behavior under the null hypothesis.  Note that polarization fractions have not been debiased.}
\label{fig:l1688}
\end{figure}

In order to test our model on real data, we used JCMT BISTRO Survey \citep{wardthompson2017} 850$\mu$m POL-2 observations of three dense clumps within the Ophiuchus L1688 molecular cloud, Oph A, Oph B and Oph C.  These data sets were originally presented by \citet{kwon2018}, \citet{soam2018}, and \citet{liu2019}, respectively, and were taken under JCMT project code M16AL004.

The Oph A and B clumps are active star formation sites, containing outflow-driving protostellar sources and gravitationally bound prestellar cores, while Oph C has little or no ongoing star formation and contains only pressure-bound cores \citep{pattle2015}.  Oph B and C are cold ($\sim 10$\,K) clumps showing little sign of external influence.  Oph A is warmer, $\sim 20$\,K, and is in the vicinity of two B stars, HD147889 (spectral type B2V, \citealt{wilking2008}; effective temperature $T_{eff}\sim20\,800$\,K, \citealt{silaj2010}) and S1 (B4V, $T_{eff}\sim17\,000$\,K; \citealt{mookerjea2018}).  The immediate proximity of the latter of these significantly increases the ionizing photon flux on Oph A over that on other clumps in L1688 \citep{pattle2015}.  The geometry and radiation field of L1688 is discussed in detail by \citet{liseau1999} and \citet{stamatellos2007}.  A finding chart for the L1688 region is shown in Figure~\ref{fig:l1688_all}.

\subsection{Observations and data reduction}
\label{sec:data_dr}

We reduced each set of observations\footnote{A single JCMT BISTRO POL-2 observation consists of 40 minutes of observing time, using the POL-2-DAISY scan pattern described by \citet{friberg2016}.} (20 observations of Oph A and C; 19 observations of Oph B) using the \emph{pol2map}\footnote{http://starlink.eao.hawaii.edu/docs/sc22.pdf} routine recently added to \textsc{Smurf} (\citealt{berry2005}; \citealt{chapin2013}) and the `January 2018' instrumental polarization model \citep{friberg2018} .  The data reduction process is as described by \citet{soam2018}, with the following modifications: (1) in the second stage of the data reduction process, the \textit{skyloop}\footnote{http://starlink.eao.hawaii.edu/docs/sc21.htx/sc21.html} routine was used, in which each iteration of the mapmaker is performed on each of the observations in the set in turn, rather than each observation being reduced consecutively as is the standard method; (2) variances in the final co-added maps were calculated according to the standard deviation of measured values in each pixel across the 20 observations, rather than as the mean of the RMS of the bolometer counts in that pixel in each observation; and (3) in the final co-added maps each observation was weighted according to the mean of its associated variance values.  The net effect of these alterations is to improve homogeneity between observations and to reduce noise in the final co-added maps.  Based on the outcomes of fitting our Monte Carlo simulations, we did not attempt to debias the measured polarization fractions. 

As discussed in Section~\ref{sec:model}, the Ricean model assumes that Stokes $Q$ and $U$ have uncorrelated uncertainties.  This is a reasonable assumption for POL-2 data.  The first step of the POL-2 data reduction process is to separate each bolometer timestream into $Q$, $U$ and $I$ timestreams by fitting a sinusoidal function to the data, using the $calcqu$\footnote{http://starlink.eao.hawaii.edu/docs/sun258.htx/sun258ss5.html} routine.  So long as the phase of the polarized emission (twice the position angle of the POL-2 half-wave plate) is accurately known at each point in the bolometer timestream, the $Q$ and $U$ timestreams should be correctly separated.  The $Q$, $U$ and $I$ data are thereafter reduced independently of one another.

In our model, we assume that the observed data can be accurately characterized by a single RMS noise value, $\sigma_{QU}$.  JCMT POL-2 observations use an observing mode wherein the central 3-arcminute-diameter region of each 12-arcminute-diameter observation has constant exposure time, and so approximately constant RMS noise \citep{friberg2016}.  We thus considered only this central region of each field.  (In Oph B, this corresponds to the Oph B2 clump.)  The three fields, along with their polarization vectors, are shown in Figure~\ref{fig:oph}.

Each pixel in the output Stokes $Q$ and $U$ maps has associated variance values $V_{Q}$ and $V_{U}$, determined as described above.  These variance maps can be converted into maps of 1-$\sigma$ RMS noise by taking their square roots.  We thus estimated a single RMS noise value for our data, 
\begin{equation}
    \langle\sigma_{QU}\rangle = \dfrac{1}{2N} \sum^{N}_{i=1}\left(\sqrt{V_{Q,i}} +\sqrt{V_{U,i}}\right),
\end{equation}
where $N$ is the number of pixels in the data set.  We set $\sigma_{QU}=\langle \sigma_{QU}\rangle$ when performing model fitting.  We retain the notation $\langle \sigma_{QU}\rangle$ while fitting real data, in order to emphasize that this is a measured property of the data set.  The $\langle\sigma_{QU}\rangle$ values measured in each field are listed in Table~\ref{tab:l1688}.  We also calculated uncertainties on polarization fraction for each pixel from the variance values $V_{Q}$, $V_{U}$ and $V_{I}$, using equation~\ref{eq:dp}.

JCMT POL-2 data are by default reduced onto 4-arcsec pixels.  In order to investigate the dependence of observed polarization fraction on RMS noise, we also gridded the data to 4$n$-arcsec pixels, where $n = 1-8$, thereby generally reducing the RMS noise, as shown in Table~\ref{tab:l1688}.  We note that binning to larger pixel sizes increases the chance of beam-averaging-related depolarization.  However, for the 8- and 12-arcsec cases, as we grid from a pixel size of $\sim 1/3$ of the 850\,$\mu$m JCMT primary beam to slightly smaller than beam-sized pixels (the JCMT has an effective beam size of 14.1 arcsec at 850$\mu$m; \citealt{dempsey2013}), we do not expect this to be an issue in at least these cases.  We did not grid to larger pixel sizes than 32 arcsec as at larger pixel sizes the RMS noise values ceased to improve, and the number of data points became prohibitively small.

\subsection{Fitting}
\label{sec:data_fitting}

We fitted each data set with the mean of the Rice distribution (equation~\ref{eq:ricemean}), using all data points and uncertainties calculated using equation~\ref{eq:dp}.  Fitting was performed as described in Section~\ref{sec:monte_fit}.  The models were normalized to the mean uncertainty in $Q$ and $U$, $\langle\sigma_{QU}\rangle$, and the data were fitted for $p_{\sigma_{QU}}$ and $\alpha$.

The results of our fitting are listed in Table~\ref{tab:l1688} and plotted in Figures~\ref{fig:l1688} and \ref{fig:32as}.  It can be seen that in all cases, the data are well-characterized by $\langle\sigma_{QU}\rangle$, and are well-described by the Rice distribution, following equation~\ref{eq:ricemean}.

Table~\ref{tab:l1688} lists the reduced $\chi^{2}$ values for each model, and for the null hypothesis.  Under the null hypothesis, $p_{\sigma_{QU}} = 0$, and so $p^{\prime} = (I/\langle\sigma_{QU}\rangle)^{-1}\sqrt{\pi/2}$ for all $I$.  As in Section~\ref{sec:monte_fit}, we note that the uncertainties on $\delta p$ can be treated as Gaussian only at high SNR.  The reduced-$\chi^{2}$ statistic is thus not strictly a direct comparison between the difference between the data and the model and the variance in the data.  We therefore compare only the relative size of the reduced-$\chi^{2}$ statistics for the best-fit and null-hypothesis models, treating a smaller value of reduced $\chi^{2}$ as being broadly representative of better agreement between the data and the model.

\begin{table*}
\centering
\begin{tabular}{@{\extracolsep{2pt}}c cc ccc@{}}
\hline
 & \multicolumn{2}{c}{Plane-of-sky distance (pc)} & \multicolumn{3}{c}{Upper-limit ionizing photon flux (s$^{-1}$m$^{-2}$)} \\ \cline{2-3} \cline{4-6}
 Region & HD147889 & S1 & HD147889 & S1 & Total \\
 \hline
 Oph A & 0.62 & 0.05 & 7.0$\times10^{10}$ & 1.5$\times10^{11}$ & 2.2$\times10^{11}$ \\
 Oph B & 1.13 & 0.51 & 2.1$\times10^{10}$ & 1.2$\times10^{9}$ & 2.2$\times10^{10}$ \\
 Oph C & 0.91 & 0.50 & 3.2$\times10^{10}$ & 1.2$\times10^{9}$ & 3.4$\times10^{10}$ \\
 \hline
 \end{tabular}
 \caption{Ionizing flux on Oph A, B and C from the stars HD147889 and S1.}
 \label{tab:fluxes}
\end{table*}

\subsection{Oph A}

Oph A shows clear evidence for a power-law behavior shallower than $\alpha=1$.  Examination of Figure~\ref{fig:l1688} shows significant deviation from $\alpha=1$ on 12-arcsec pixels.  This is confirmed by the fitted models producing a reduced-$\chi^{2}$ statistic approximately $3 - 4$ times smaller than that of the null hypothesis on all pixel sizes.

When fitting the mean of the Rice distribution to the Oph A data, we consistently recovered $p_{100\,mJy/beam} \approx 0.047$ and $\alpha \approx 0.34$.  The results returned from 4-arcsec to 28-arcsec pixel data are consistent within fitting uncertainties, with no obvious signs of depolarization due to beam averaging, although the fitting results become more uncertain as pixel size increases, as shown in Figure~\ref{fig:l1688_pix}.  Fitting of the Oph A data fails for 32-arcsec pixels: this is likely due to the relatively small number of remaining data points, and the significant intrinsic scatter in the data around the best-fit model.  In all cases, the best-fitting models produce reduced-$\chi^{2}$ statistics significantly greater than unity, suggesting more variation in the data than can be explained by our simple model alone.

We thus interpret our results as indicating that the Oph A data can be represented by the model $p_{100{\rm mJy/beam}}=0.047$ and $\alpha=0.34$. Equations~\ref{eq:p100} and \ref{eq:sn} suggest that, when $\sigma_{QU}\approx 1$\,mJy/beam, this behavior will be recoverable only when a significant number of data points fall above $(I/\sigma_{QU})_{crit}\sim 18$.

\subsection{Oph B}

In Oph B, the incompatibility of the data with the null hypothesis becomes more apparent with increasing pixel size.  In the 4-arcsec case, the Oph B data do not clearly support a power-law index distinct from $\alpha = 1$ being observed, with the Ricean-mean model and the null hypothesis producing similar reduced-$\chi^{2}$ values.

\begin{figure}
\centering
\includegraphics[width=0.49\textwidth]{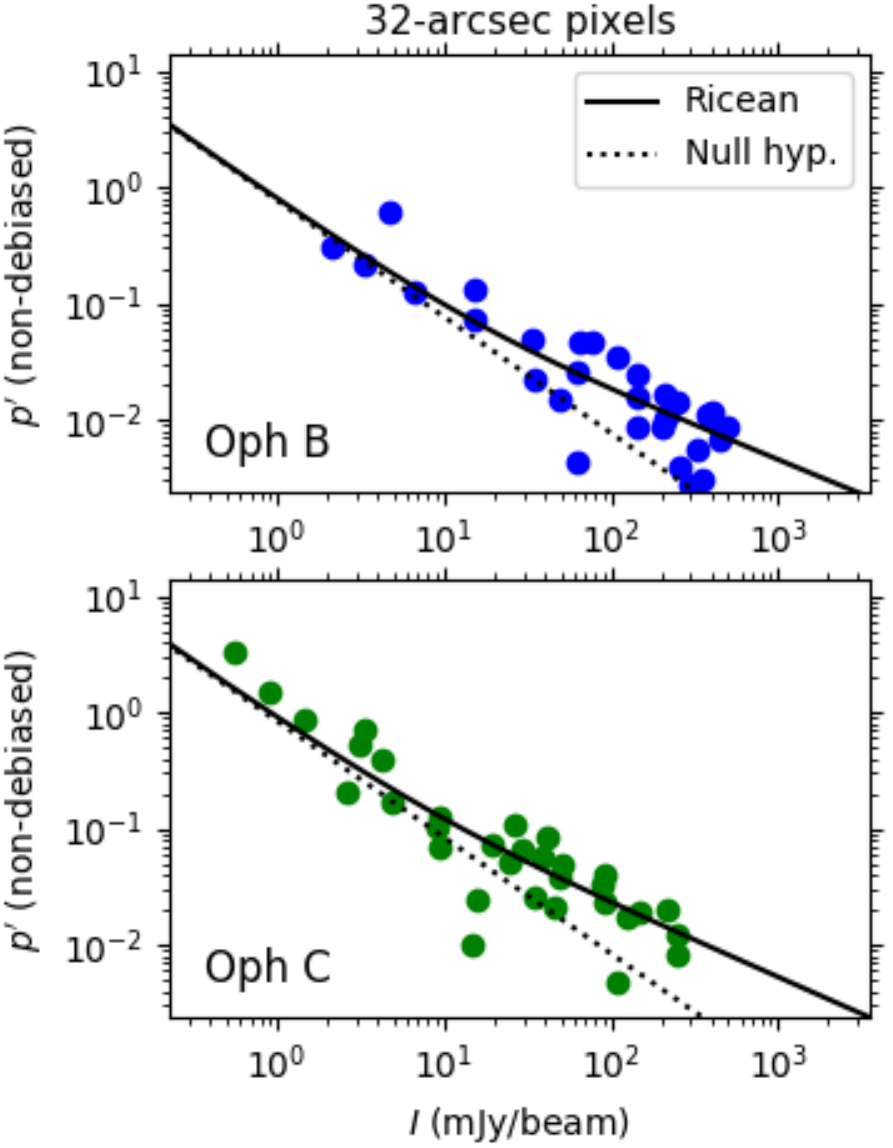}
\caption{Observed polarization fraction $p^{\prime}$ versus total intensity $I$ for the B and C data, with data gridded to 32-arcsec pixels.  Polarization fractions have not been debiased.  Best-fitting models are shown as in Figure~\ref{fig:l1688}.  Note that both data sets show significant deviation from the null-hypothesis behavior.}
\label{fig:32as}
\end{figure}

\begin{figure*}
\centering
\includegraphics[width=\textwidth]{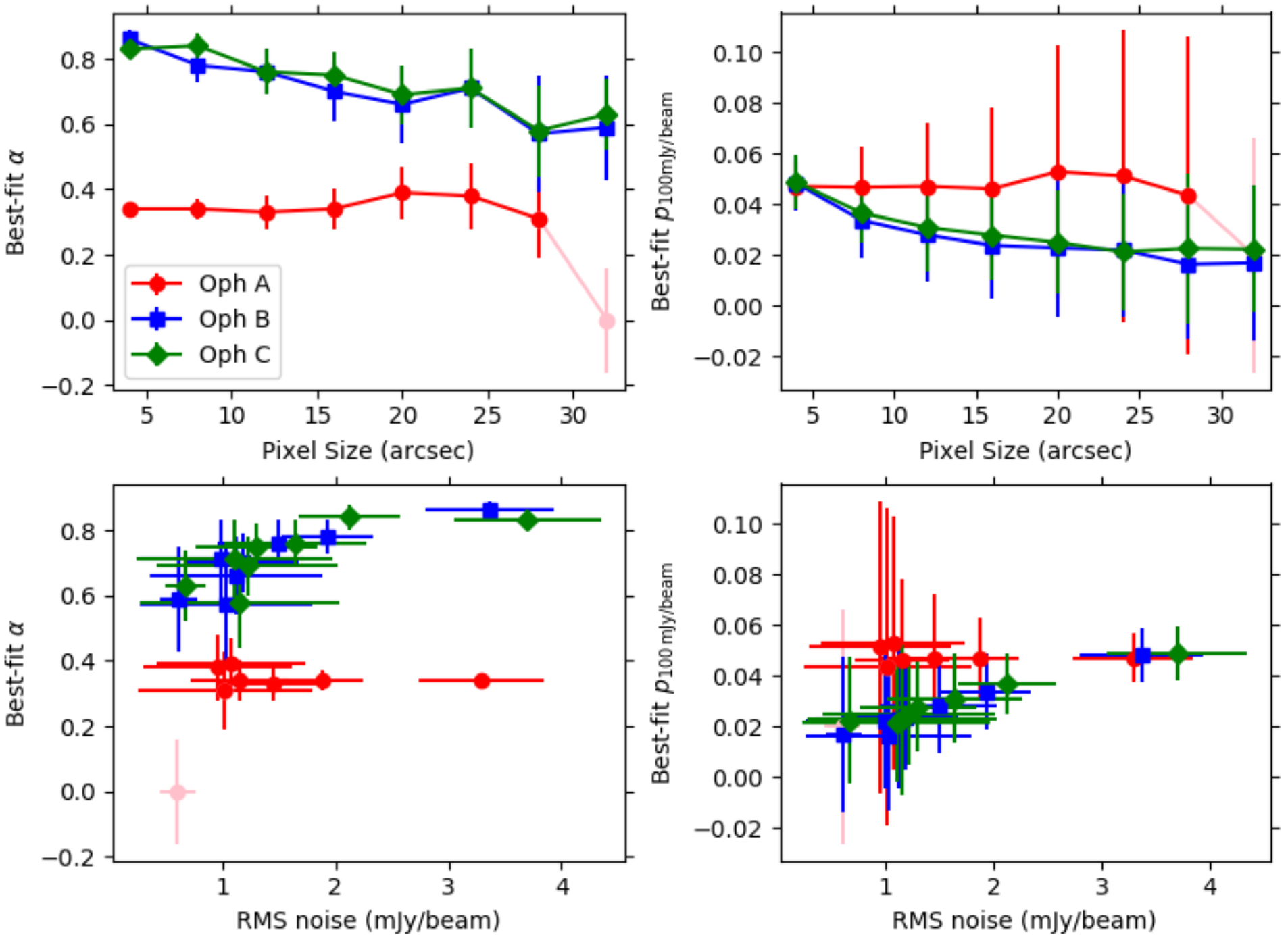}
\caption{Best-fit values of $\alpha$ (left) and $p_{100{\rm mJy/beam}}$ (right) for the Oph A, B and C data, as determined from fitting the Ricean-mean model, as a function of pixel size (top) and of RMS noise $\langle\sigma_{QU}\rangle$ (bottom).  Fitting results for Oph A on 32-arcsec pixels are shown in pink as this result is not reliable.}
\label{fig:l1688_pix}
\end{figure*}

In the 12-arcsec case, the data appear to be skewed above the null hypothesis line in Figure~\ref{fig:l1688}.  The Ricean-mean fitting results produce reduced-$\chi^{2}$ values a factor $\sim 2$ smaller than that of the null hypothesis.  Gridding the data to larger pixel sizes results in progressively smaller values of $p_{\sigma_{QU}}$ and $\alpha$ being recovered, as shown in Figure~\ref{fig:l1688_pix}.  The 32-arcsec case is shown in Figure~\ref{fig:32as}, and clearly shows that the data are not consistent with an index of $\alpha = 1$.

The most representative values of $\alpha$ and $p_{100{\rm mJy/beam}}$ in Oph B are not very well-constrained.  To a certain extent, Figure~\ref{fig:l1688_pix} shows a stabilization in the fitted values of $\alpha$ and $p_{100{\rm mJy/beam}}$ in Oph B at lower RMS noise values.  On the larger pixel sizes considered, $\alpha \sim 0.6-0.7$. On 20-arcsec pixels and larger, $p_{100{\rm mJy/beam}}$ becomes consistent with a value $\sim 0.02$.  It is also possible that gridding to larger pixel sizes alters the observed average grain properties in Oph B, although, as discussed above, this does not appear to be the case in Oph A.  Equations~\ref{eq:p100} and \ref{eq:sn} suggest that $\alpha = 0.65$ and $p_{100{\rm mJy/beam}}=0.02$ would, for $\sigma_{QU}\approx 1$\,mJy/beam, be recoverable only when a significant number of data points fall above $(I/\sigma_{QU})_{crit}\sim 26$.

\subsection{Oph C}

We find that Oph C behaves similarly to Oph B.  Due to its lower peak brightness than Oph B, the Oph C data show little evidence for deviation from an $I^{-1}$ behavior on beam-sized or smaller pixels.  In the 4- and 8-arcsec data the null hypothesis produces a reduced-$\chi^{2}$ value comparable to that of the Ricean-mean model, and there is no clear evidence that the Ricean-mean model provides a better description of the data.  In the 12-arcsec case, shown in Figure~\ref{fig:l1688}, there is only a marginal improvement in goodness-of-fit over the null-hypothesis case.

As in Oph B, gridding to larger pixels produces smaller values of both $\alpha$ and $p_{100{\rm mJy/beam}}$, and makes the deviation of the data from the null hypothesis behavior more apparent.  Figure~\ref{fig:32as} qualitatively shows the similarity between the Oph B and Oph C data, while Figure~\ref{fig:l1688_pix} shows that the fitting results of Oph B and C agree very closely in all cases.

\subsection{Discussion}
\label{sec:data_discussion}

Our results suggest that grains in Oph A are intrinsically better aligned with the magnetic field than those in Oph B and Oph C.  Grains in Oph B and C appear to lose what alignment they have with the magnetic field more precipitously with increasing density (or extinction) than is the case in Oph A.  Oph B and C appear to have indistinguishable grain properties despite their differing star formation histories, while Oph A and B behave significantly differently despite having comparable mass and both being active sites of star formation.

We estimated the upper-limit ionizing fluxes on Oph A, B and C from the B stars HD147889 and S1 as a qualitative indicator of the differences in interstellar radiation field (ISRF) on the three clumps.  We took the flux of Lyman continuum photons to be $\sim 10^{20.4}$\,cm$^{-2}$\,s$^{-1}$ from the surface of HD147889 and $\sim 10^{18.5}$\,cm$^{-2}$\,s$^{-1}$ from the surface of S1 (\citealt{pattle2015}, and refs. therein).  We determined plane-of-sky distances from HD147889 and S1 to the centers of Oph A, B and C assuming a distance to L1688 of 138\,pc \citep{ortizleon2018}, and so estimated upper-limit ionizing fluxes from the stars on each region\footnote{We note that if we adopted the three-dimensional model of L1688 proposed by \citet{liseau1999}, the distance from HD147889 to the three clumps would be: 0.86\,pc to Oph A, 1.33\,pc to Oph B, and 1.03\,pc to Oph C (distances scaled to account for their assumed distance to L1688 of 150\,pc).  This would alter our inferred fluxes from HD147889 by factors $\sim 0.5$, $\sim 0.6$ and $\sim 0.8$ in Oph A, B and C respectively.  This would not change the conclusions of our analysis.}.  These fluxes are listed in Table~\ref{tab:fluxes}.  We assume that the global ISRF on the three regions is comparable (a reasonable assumption given the clumps' proximity to one another).  It can be seen that the ionizing flux on Oph A is an order of magnitude larger than that on Oph B and Oph C, and that this difference is primarily caused by the proximity of S1 to Oph A.  HD147889 is likely to affect the three clumps similarly, while S1's influence is dominant in Oph A but negligible elsewhere.

The most likely explanation for the better grain alignment in Oph A is thus the elevated photon flux on that region, primarily resulting from the proximity of the star S1.  Under the radiative torque alignment (RAT) paradigm of grain alignment (\citealt{lazarian2007}; \citealt{andersson2015}), this stronger and bluer radiation field on Oph A would allow grain alignment to persist to higher optical depth.  The strongly anisotropic radiation field on Oph A might also favor better grain alignment in this region \citep{dolginov1976,onaka2000,weingartner2003}.  We note that the difference in behavior between Oph A and Oph B, both of which are actively forming stars, suggests that the better grain alignment in Oph A is primarily driven by external influence, and not by short-wavelength flux from protostellar sources within the clump.

We note that the ionizing flux from these stars will not itself directly contribute to grain alignment within the clumps.  In order to drive RAT grain alignment, the wavelength of the incident radiation must be shorter than twice the size of the largest grains, i.e. $\lesssim1-2\,\mu$m \citep{andersson2015}.  The short-wavelength photons considered here will undergo multiple scatterings before they can contribute significantly to grain alignment, while longer-wavelength emission from the stars may contribute more directly, particularly in Oph A, thanks to its proximity to S1.  We emphasize that the calculations above only qualitatively demonstrate the global elevation of photon flux on Oph A over the other two clumps.  Modeling of the detailed radiation field in L1688 is beyond the scope of this work. 

The magnetic field in Oph A may also be intrinsically more ordered in Oph A than in Oph B and C, as the clump's location between HD147889 and S1 could result in its molecular gas, and so its magnetic field, being compressed by the HD147889 photon-dominated region (PDR) and the S1 reflection nebula.  In contrast, Oph B and C are evolving in relative isolation from the two B stars, and are not undergoing significant compression.  In this case, the lower value of $\alpha$ in Oph A might in part result from its more ordered internal magnetic field, with less vector cancellation of observed polarization fraction along the line of sight occurring in Oph A than in the other two regions.

Another possible cause of better grain alignment in Oph A than in the other two clumps is grain growth in the dense regions of Oph A.  The peak gas density of Oph A is approximately one order of magnitude higher than in Oph B and C \citep{motte1998}.  Such high densities might provide the necessary conditions for the formation of large dust grains (e.g. \citealt{hirashita2013}).  In the RAT paradigm, larger dust grains can be aligned by longer-wavelength photons, as described above, and thus the presence of large grains would allow grain alignment to persist to higher optical depth.

While our results support better grain alignment in Oph A than in the other clumps, they do not suggest that grains in Oph B and C have no alignment with the magnetic field.  Our modeling suggests that an index $\alpha \sim 0.6-0.7$ is plausible for both Oph B and Oph C, suggesting that some degree of grain alignment may persist to high optical depths within these clumps.

These results suggest that grain alignment could persist to significantly higher densities within starless clumps and cores than has previously been believed to be the case (e.g \citealt{jones2015}; \citealt{kwon2018}, \citealt{soam2018}) even in the absence of a short-wavelength illuminating source.  This is consistent with recent modelling results, which suggest that grains remain well-aligned with the magnetic field at gas densities $>10^{3}$\,cm$^{-3}$ \citep{seifried2019}.

\subsection{Limitations of the fitting process}
\label{sec:data_limitations}

The simple model which we consider in this work is subject to a number of limitations.

We emphasize that in this work we selected our data such that they can be well characterized by a single RMS noise value in both Stokes $Q$ and $U$.  Data sets containing significant variation in RMS noise would produce additional vertical spread in the $p^{\prime}-I$ plane, further complicating the recovery of an accurate value of $\alpha$.

Our data show scatter about the best fit line greater than can be explained by instrumental uncertainty alone, particularly in Oph A, which in all cases shows reduced-$\chi^{2}$ values significantly larger than those in Oph B and C, and where fitting fails for the largest pixel size considered here, likely due to significant intrinsic scatter in the data and the small number of data points to which the model can be fitted.  In order to demonstrate the statistical properties of the data, we have chosen a very simple model in which the data are characterized by a single power law.  More complex relationships between polarization fraction and intensity could be investigated in future studies, as well as accounting for intrinsic variation in $p_{\sigma_{QU}}$ and $\alpha$ within a given region.

Our model is unphysical in that it suggests that polarization fraction can increase indefinitely at small $I$.  We are implicitly assuming firstly that there is a turnover in behavior at a density below which polarization fraction becomes a shallower function of intensity, tending to the value in the low-density ISM, and secondly that this turnover occurs at densities lower than we can probe with POL-2.  (In approximately isothermal environments, the SCUBA-2 camera is effectively volume-density-limited in its detections - see \citet{wardthompson2016}.)  

Our results suggest that an observed index of $\alpha \simeq 1$ in submillimeter data (e.g. \citealt{alves2014}; \citealt{jones2015}, \citealt{liu2019}) is not sufficient to claim that non-aligned grains have been observed.  However, if a break or turnover in behavior from a shallow power law ($\alpha<1$) to an index of $\alpha=1$ with \emph{increasing} intensity were observed within a single polarized submillimeter emission data set, this would be strong evidence for loss of grain alignment at high intensities.  This is because if an index $\alpha<1$ were recoverable at intermediate intensities, then an $\alpha=1$ index seen at high intensities in the same data set could then not simply result from having insufficient signal-to-noise to measure a shallower index, and would thus be indicative of a genuine change in behaviour with intensity.  We do not see any evidence of such a break in behavior in Ophiuchus.

\subsection{Relation between $I$ and visual extinction $A_{V}$}
\label{sec:av}

In this paper we consider only the relationship between $p$ and $I$, with the goal of accurately determining their underlying relationship in the presence of Ricean noise.  However, it is important to emphasize that the true physical relationship under investigation is not between $p$ and $I$ but between $p$ and $A_{V}$.

Recent POL-2 studies (\citealt{juvela2018}; \citealt{wang2019}; \citealt{coude2019}) show a shallower relationship between $p^{\prime}$ and $Herschel$ Space Observatory-derived dust opacity/optical depth measurements (a proxy for $A_{V}$, as discussed below) than between $p^{\prime}$ and $I$.  We here present a simple argument for why we expect the $p-A_{V}$ relationship to be intrinsically shallower than the $p-I$ relationship.

The submillimeter intensity $I_\nu$ of thermal dust emission at frequency $\nu$ is given by
\begin{equation}
    I_{\nu} = B_{\nu}(T)\tau_{\nu} \Omega,
\end{equation}
where $B_{\nu}(T)$ is the Planck function at temperature $T$, $\tau_{\nu}$ is submillimeter optical depth, and $\Omega$ is solid angle (e.g \citealt{hildebrand1983}).  We henceforth assume that we are observing over a constant area, and so
\begin{equation}
    I_{\nu} \propto B_{\nu}(T)\tau_{\nu}.
    \label{eq:ibtau}
\end{equation}
Assuming $\tau_{\nu} \propto A_{V}$, i.e. optically thin submillimeter emission (e.g. \citealt{jones2015}) and constant dust optical properties along the line of sight,
\begin{equation}
    I_{\nu} \propto B_{\nu}(T)A_{V}.
    \label{eq:ita}
\end{equation}
$I$ is thus a direct tracer of $A_{V}$ only where $T$ is constant.  However, in most environments in molecular clouds, $B_{\nu}(T)$ and $I$ are observed to be anti-correlated (e.g. \citealt{kirk2013}; \citealt{konyves2015}).  This is a physical effect, with cooling being caused by self-shielding in dense environments (e.g. \citealt{glover2012}).  Such cooling is expected in regions which do not contain embedded massive stars causing internal heating, such as the dense clumps of L1688 \citep{stamatellos2007}.  We parameterize the relationship between $B_{\nu}(T)$ and $I_{\nu}$ as
\begin{equation}
         B_{\nu}(T) \propto I_{\nu}^{\gamma},
    \label{eq:TI}
\end{equation} \black
initially placing no constraint on $\gamma$ (note that here $B_{\nu}(T)$ is the source function of the dust emission; c.f. equation~\ref{eq:ibtau}).  Combining equations \ref{eq:p_model}, \ref{eq:ita} and \ref{eq:TI},
\begin{equation}
    p\propto A_{V}^{-\frac{\alpha}{1-\gamma}}.
\end{equation}
If $\gamma<0$ ($B_{\nu}(T)$, and so $T$, decreases with increasing $I_{\nu}$), then
\begin{equation}
    \dfrac{1}{1-\gamma} < 1.
\end{equation}
Thus, in most physical environments in molecular clouds, any power-law relationship between $A_{V}$ and $p$ must be shallower (and likely more weakly correlated) than that between $I$ and $p$.   We note that $\gamma<0$ would not hold in the presence of significant heating by sources located at high $A_{V}$, but in that case, we might expect these heating sources to also be driving grain alignment in their vicinity.

This analysis further suggests that grains could better aligned at higher extinction than has previously been believed.

The exact nature of the relationship between $I$, $T$ and $\tau$ is not obtainable from single-wavelength observations such as those considered here.  However, forthcoming multi-wavelength studies will allow more direct investigation of the $I-A_{V}$ relationship.

\section{Summary} \label{sec:summary}

The dependence of polarization fraction on total intensity in polarized submillimeter emission measurements is typically parameterized as a power law, and used to infer the efficiency of dust grain alignment with the magnetic field in star-forming clouds and cores.  In this work we have demonstrated that significant signal-to-noise and well-characterized noise properties are required to recover a genuine power-law relationship between polarization fraction and total intensity.

We presented a simple model for the dependence of polarization fraction on total intensity in molecular clouds, and so demonstrated that below a signal-to-noise threshold of $I/\sigma_{QU}=(p_{\sigma_{QU}}^{-1}\sqrt{\pi/2})^{\frac{1}{1-\alpha}}$, a power-law index of $-1$ will \emph{always} be observed, as a result of the addition in quadrature of Stokes $Q$ and $U$ components, and of the approximate $1/I$ dependence of error in polarization fraction.  For power-law indices $\alpha <1$, the intrinsic dependence of polarization fraction on intensity will be recoverable at high signal-to-noise.  However, a genuine measurement of $p\propto I^{-1}$ -- indicating un-aligned grains -- will be indistinguishable from statistical noise in most if not all physically realistic scenarios without additional information. We demonstrated that fitting a single power law is likely to result in overestimation of $\alpha$, and so of the degree of depolarization occurring.  We further found that fitting the mean of the Rice distribution to non-debiased data will accurately recover both $p_{\sigma_{QU}}$ and $\alpha$, provided a reasonable number of data points fall above the required signal-to-noise threshold.

We used JCMT POL-2 observations of three clumps in the L1688 region of the Ophiuchus molecular cloud to demonstrate the statistical behavior described above.  We found that the Oph A region, which is illuminated by two B stars, shows significantly better grain alignment than the neighboring Oph B and C.  We found a power-law index of $\alpha \approx 0.34$ in Oph A, significantly shallower than found by previous works.  The power-law indices in Oph B and C are less well-constrained, but are steeper than that of Oph A, and are likely to be in the range $\alpha \sim 0.6-0.7$.  Oph B and Oph C have intrinsically lower polarization fractions than Oph A at a total intensity of 100\,mJy/beam, with emission from Oph A being 4.7\% polarized, while emission from Oph B and C is $\sim 2$\% polarized.  Oph C, a quiescent cloud, appears to behave comparably to the actively star-forming Oph B.  Our results thus suggest that grain alignment in Ophiuchus is driven by the external radiation field on the clumps, and not by internal radiation sources.

These results suggest that grain alignment could persist to significantly higher densities within starless clumps and cores than has previously been believed to be the case.  Submillimeter polarization measurements could thus potentially trace the magnetic field morphology in dense, star-forming gas.

\acknowledgments

K.P., S.P.L. and J.W.W. acknowledge support from the Ministry of Science and Technology (Taiwan) under Grant No. 106-2119-M-007-021-MY3. K.P. was an International Research Fellow of the Japan Society for the Promotion of Science for part of the duration of this project.  W.K. and C.W.L were supported by Basic Science Research Program through the National Research Foundation of Korea (NRF-2016R1C1B2013642 and 2019R1A2C1010851, respectively).  The James Clerk Maxwell Telescope is operated by the East Asian Observatory on behalf of The National Astronomical Observatory of Japan; Academia Sinica Institute of Astronomy and Astrophysics; the Korea Astronomy and Space Science Institute; Center for Astronomical Mega-Science (as well as the National Key R\&D Program of China with No. 2017YFA0402700). Additional funding support is provided by the Science and Technology Facilities Council of the United Kingdom and participating universities in the United Kingdom and Canada.  The authors wish to recognize and acknowledge the very significant cultural role and reverence that the summit of Maunakea has always had within the indigenous Hawaiian community. We are most fortunate to have the opportunity to conduct observations from this mountain.

\vspace{5mm}
\facilities{James Clerk Maxwell Telescope (JCMT)}

\software{Starlink \citep{currie2014}, Astropy \citep{astropy, astropy2018}}

\section*{Appendix: List of symbols used in this work}

\FloatBarrier

\begin{table*}
    \centering
    \begin{tabular}{cccc}
    \hline
    Symbol & Definition & Units & Section defined in\\
    \hline
    $\alpha$ & Power-law index, $p\propto I^{-\alpha}$ & Dimensionless & \ref{sec:intro}, \ref{sec:model} \\
    $\alpha_{max}$ & Steepest recoverable $\alpha$ for given $p_{\sigma_{QU}}$ and $(I/\sigma_{QU})_{max}$ & Dimensionless & \ref{sec:monte_fit} \\
    \rule{0pt}{0ex} $I$ & Stokes $I$ intensity & Intensity & \ref{sec:equations} \\
    $\sigma_{I}$ & RMS noise in Stokes $I$ & Intensity & \ref{sec:equations} \\ 
    $\delta I$ & Measurement uncertainty on Stokes $I$ & Intensity & \ref{sec:equations} \\
    $I_{obs}$ & Stokes $I$ intensity in Monte Carlo model & Intensity & \ref{sec:monte} \\
    \rule{0pt}{0ex} $\mathcal{I}_{0}$ & Modified Bessel function of order 0 & -- & \ref{sec:equations_rice} \\
    $\mathcal{I}_{1}$ & Modified Bessel function of order 1 & -- & \ref{sec:equations_rice} \\
    $\mathcal{L}_{\frac{1}{2}}$ & Laguerre polynomial of order $\frac{1}{2}$ & -- & \ref{sec:equations_rice} \\
    \rule{0pt}{0ex} $N$ & Number of pixels in a given data set & Dimensionless & \ref{sec:data_dr} \\
    \rule{0pt}{0ex} $p$ & Intrinsic polarization fraction & Dimensionless & \ref{sec:equations} \\
    $p^{\prime}$ & Measured polarization fraction & Dimensionless & \ref{sec:equations_rice}; see also \ref{sec:monte_fit} \\
    $p^{\prime}_{db}$ & Debiased measured polarization fraction & Dimensionless & \ref{sec:equations_debias} \\
    $p_{0}$ & Polarization fraction at reference intensity $I_{0}$ & Dimensionless & \ref{sec:model} \\
    $p_{\sigma_{QU}}$ & Polarization fraction at reference intensity $\sigma_{QU}$ & Dimensionless & \ref{sec:model} \\
    $p_{100{\rm mJy/beam}}$ & Polarization fraction at $I=100$\,mJy/beam & Dimensionless & \ref{sec:model} \\
    $\sigma_{p}$ & RMS noise in polarization fraction & Dimensionless & \ref{sec:equations_rice} \\
    $\delta p$ & Measurement uncertainty on polarization fraction & Dimensionless & \ref{sec:empiric_uncert} \\
    $\mu_{p}$ & Mean of Rice-distributed polarization fraction & Dimensionless & \ref{sec:equations_rice} \\
    \rule{0pt}{0ex} $P$ & Intrinsic polarized intensity & Intensity & \ref{sec:equations} \\
    $P^{\prime}$ & Measured polarized intensity & Intensity & \ref{sec:equations_rice} \\
    $P^{\prime}_{db}$ & Debiased measured polarized intensity & Intensity & \ref{sec:equations_debias} \\
    \rule{0pt}{0ex} $Q$ & Stokes $Q$ intensity & Intensity & \ref{sec:equations} \\
    $\sigma_{Q}$ & RMS noise in Stokes $Q$ & Intensity & \ref{sec:equations} \\
    $\delta Q$  & Measurement uncertainty on Stokes $Q$ & Intensity & \ref{sec:equations} \\
    $V_{Q}$ & Variance on Stokes $Q$ intensity & (Intensity)$^{2}$ & \ref{sec:data_dr} \\
    $Q_{obs}$ & Stokes $Q$ intensity in Monte Carlo model & Intensity & \ref{sec:monte} \\
    \rule{0pt}{4ex} $U$ & Stokes $U$ intensity & Intensity & \ref{sec:equations}\\
    $\sigma_{U}$ & RMS noise in Stokes $U$ & Intensity  & \ref{sec:equations} \\ 
    $\delta U$ & Measurement uncertainty on Stokes $U$ & Intensity & \ref{sec:equations} \\
    $V_{U}$ & Variance on Stokes $U$ intensity & (Intensity)$^{2}$ & \ref{sec:data_dr} \\
    $U_{obs}$ & Stokes $U$ intensity in Monte Carlo model & Intensity & \ref{sec:monte} \\
    \rule{0pt}{0ex} \multirow{2}{*}{$\sigma_{QU}$} & \multirow{2}{*}{\shortstack[l]{(1) Representative RMS noise in Stokes Q and U \\ (2) Reference intensity for fitted model}} & \multirow{2}{*}{Intensity} & \multirow{2}{*}{\ref{sec:model}}\\
     &  & \\
    \rule{0pt}{0ex} $\langle\sigma_{QU}\rangle$ & $\sigma_{QU}$ inferred from real data & Intensity & \ref{sec:data_dr} \\
    \rule{0pt}{0ex} $\theta_{p}$ & Polarization angle & Angle & \ref{sec:equations_angle} \\
    $V$ & Stokes $V$ intensity & Intensity & \ref{sec:equations} \\
    \rule{0pt}{0ex} $(I/\sigma_{QU})_{crit}$ & Critical SNR below which $p^{\prime}\propto I^{-1}$ dominates & Dimensionless & \ref{sec:model_SNcrit}\\
    $(I/\sigma_{QU})_{max}$ & Maximum SNR in a given data set & Dimensionless & \ref{sec:monte_fit}\\
    \rule{0pt}{0ex} $A_{V}$ & Visual extinction & Magnitudes & \ref{sec:av} \\
    $B_{\nu}$ & Planck function & W\,m$^{-2}$\,Hz$^{-1}$\,sr$^{-1}$ & \ref{sec:av} \\
    $\gamma$ & Power-law index, $I\propto T^{-\gamma}$ & Dimensionless & \ref{sec:av} \\    
    $\Omega$ & Solid angle & Steradians & \ref{sec:av} \\
    $\nu$ & Frequency & Hertz & \ref{sec:av} \\
    $T$ & Dust temperature & Kelvin & \ref{sec:av} \\
    $\tau$ & Submillimeter optical depth/dust opacity & Dimensionless & \ref{sec:av} \\
    \hline
    \end{tabular}
    \caption{Symbols used in this work}
    \label{tab:symbols}
\end{table*}

\end{document}